\documentclass{cup-elements}

\usepackage{blindtext}
\usepackage{multirow}
\usepackage{graphicx}
\usepackage[natbibapa]{apacite}

\CUPseries{Elements in the Structure and Dynamics of Complex Networks}
\CUPelements{Boolean Networks as Predictive Models}

\title{Boolean Networks as Predictive Models of Emergent Biological Behaviors}

\author{Jordan C. Rozum}
\affil{Binghamton University (State University of New York)}

\author{Colin Campbell}
\affil{University of Mount Union}

\author{Eli Newby}
\affil{Pennsylvania State University}

\author{Fatemeh Sadat Fatemi Nasrollahi}
\affil{Indiana University}

\author{R{\'e}ka Albert}
\affil{Pennsylvania State University}

\begin{document}

\frontmatter  
\maketitle

\begin{abstract}
 Interacting biological systems at all organizational levels display emergent behavior. Modeling these systems is made challenging by the number and variety of biological components and interactions (from molecules in gene regulatory networks to species in ecological networks) and the often-incomplete state of system knowledge (e.g., the unknown values of kinetic parameters for biochemical reactions). Boolean networks have emerged as a powerful tool for modeling these systems. We provide a methodological overview of Boolean network models of biological systems. After a brief introduction, we describe the process of building, analyzing, and validating a Boolean model. We then present the use of the model to make predictions about the system's response to perturbations and about how to control (or at least influence) its behavior. We emphasize the interplay between structural and dynamical properties of Boolean networks and illustrate them in three case studies from disparate levels of biological organization.  
 
\end{abstract}

\keywords{Boolean networks; complex networks; signal transduction; gene regulatory network; mutualistic network; discrete dynamics; state transition graph; attractor identification; trap sets; stable motifs; attractor control}

\JEL{A12, B34, C56, D78, E90}

\copyrightauthor{Jordan C. Rozum, Colin Campbell, Eli Newby, Fatemeh Sadat Fatemi Nasrollahi, R{\'e}ka Albert, 2023}

\mainmatter  

\section{Types of biological networks}

Interacting systems abound at every level of biological organization, from the molecular to the population level. For example, molecular interacting systems consist of genes, their transcripts (mRNAs), proteins and small molecules; their interactions include gene transcription, protein translation, protein-protein interactions, and chemical reactions. At the other extreme, interactions among  animals include predation, competition, and symbiosis.

A fundamental goal of biology is to understand why biological systems behave the way they do. One promising avenue toward this goal is to study how interacting biological systems at each level determine the emergent behavior at the next level. Emergence has been defined and studied in various contexts \citep{hopfield_neural_1982,crutchfield_calculi_1994,koorehdavoudi_statistical_2016,kivelson_defining_2016,zheng_introduction_2021}; in this review, emergence refers to the collective behaviors of system components that arise from their interactions in a nontrivial way.  For example, cellular decisions and phenotypes arise from the interactions of numerous molecular components. Similarly, interactions among cells determine how multi-cellular organisms develop and how tissues and organs function; interactions among individuals form the basis of social communities; and interactions among species underlie ecological communities. However, not every interaction plays the same role in determining the behavior of the system as a whole.

A network representation of a system of coupled elements can be an effective tool for studying how interactions give rise to collective behavior. A network (or graph) is a mathematical abstraction, consisting of nodes, which represent different elements, and edges, which specify the pairwise relationships between the elements. For example, in molecular networks, nodes represent genes, RNA, proteins, and small molecules; edges indicate reactions, interactions, and regulatory relationships. The edges of a network can be symmetrical (representing a mutual relationship) or directed (representing mass or information flow from a source to a target). Edges can also have a sign, representing positive (activating) or negative (inhibitory) influences and relationships. For example, a possible network representation of interactions among the members of microbial communities would use directed and signed edges in which the sign of each edge indicates whether the interaction is beneficial or detrimental to the survival of the microbe to which the edge directs. A network representation allows the use of graph measures to characterize the organization of the interactions in biological systems, but does not fully capture the system's higher-level emergent properties. A second important step toward the elucidation of higher-level emergent properties is the modeling of the propagation of information in these networks \citep{ramo_measures_2007, barros_information_2009, harush_dynamic_2017, maheshwari_2017}.


\section{Modeling the dynamics of biological networks} 

Each element of a biological interacting system (represented as a node in a biological network) can be characterized by an abundance or activity. For example, a molecular species is characterized by an abundance (respectively measured as a concentration or population); an ion channel is represented by an activity (open or closed). As each abundance or activity can change in time due to the interactions in the system, we need additional dynamical information to supplement the static network representation and capture the higher-level emergent properties of the system. This information is provided as a dynamic model.

In a dynamic model on a network, each node, $i$, is assigned a state variable $X_i$ to represent its abundance or activity. The time variation of $X_i$ is determined by a regulatory function for this variable, $f_i$, whose inputs correspond to the regulators of the node $i$  in the network. In this way, information (realized as a state change of a node) propagates through the network. Each node's state will evolve over time, eventually converging into a long-term behavior such as a steady state or a sustained oscillation. The emergent behaviors of the system (e.g., the phenotypes of a gene regulatory model) can then be characterized by the set of possible long-term behaviors of all nodes, or of a relevant subset of nodes. Dynamical models can be classified into continuous or discrete models depending on whether the state variables are continuous or discrete. Choosing the right dynamical modeling framework involves striking a balance between modeling detail and scalability to large system sizes. Other considerations for choosing a modeling framework include the specifics of the system being modeled and the intended use of the model, as we will discuss in Section \ref{model_features}.

\subsection{Continuous modeling}

 There are many types of continuous dynamical models, but ordinary differential equations (ODEs) are by far the most common. In ODE models on networks, the rate of change (time derivative) of each node state $X_{i}(t)$ is expressed as a function of variables in the biological network. The positive influences and regulatory effects incorporated in the network contribute to positive terms in $\frac{dX_{i}}{dt}$, and the negative influences and regulatory effects contribute to negative terms or decrease the magnitude of positive terms. Continuous models of molecular networks usually include a negative term that corresponds to spontaneous decay of the molecule (due, for example, to dilution effects during cell division); such a term is usually not explicitly represented as a self-edge in the network. ODEs are suited for well-characterized systems, where the mechanistic details of each interaction are known through collecting a sufficient amount of quantitative information (usually through decades of experimental work). Elementary biochemical reactions are usually described using mass action kinetics, in which the rate of a chemical reaction is proportional to the concentration of the reactants. Other often used functional forms are Michaelis-Menten kinetics, according to which the rate of an enzyme-catalyzed reaction depends on the concentration of the reactant as $\frac{R}{K_M+R}$, and Hill functions, which generalize the latter to $\frac{R^h}{K_M^h+R^h}$. Each of these functions requires extensive parameterization, which needs data of good quality and quantity. The challenges of parameter estimation for systems with known regulatory functions are reviewed by \cite{loskot_comprehensive_2019}.  Due to insufficient experimental data, the state of knowledge in many biological systems is  incomplete to an extent that makes continuous modeling very challenging: in these systems not all interactions have been established, and even for known causal relationships the underlying mechanisms are not known. 
 
 This uncertainty poses a challenge for continuous modeling of these systems. To tackle these challenges in biomolecular systems, modelers often exploit parameter robustness in the ODEs of interest. In these systems, long-term behaviors often do not depend very strongly on the precise value of reaction rates or other kinetic parameters; e.g., this is a well-known property of the \emph{Drosophila} segment polarity network \citep{von_dassow_segment_2000,von_dassow_design_2002}. This robustness is closely related to the ways in which the ODEs can be discretized, which has been explored by \cite{thomas_multistationarity_2001,thomas_multistationarity_2001-1}. More recently, we have examined how the connection between discrete and continuous models can be exploited in these systems \citep{rozum_identifying_2018,rozum_self-sustaining_2018,rozum_controlling_2019}. Modelers have also found success in dealing with the challenges of biological uncertainty by exploiting dynamical constraints imposed by the network of interactions for controlling ODEs, as reviewed by \cite{liu_control_2016,rozum_leveraging_2022}. For an introduction to continuous modeling in biology, we refer the reader to \cite{mackey_simple_2016}.

\subsection{Discrete modeling}
Discrete dynamical models use discrete variables to represent logic categories of node abundance and describe the future state of each node as a function of the current states of its regulators. Discrete models only require qualitative or relative measurements (for example, information that a protein is much more abundant in one environmental condition compared to a different condition). As a result, they use no or few kinetic parameters, and can generally handle larger networks than continuous models. Systems in which each variable has a multi-modal distribution are best suited for discrete dynamic modeling. The outputs of discrete models can be made more continuous by performing ensembles of simulations that differ in the initial condition or in the order of events (akin to simulating a population of cells) and reporting ensemble averages of each variable. Discrete dynamic models have been used successfully to describe the attractor repertoire of a variety of biological systems, as reviewed by \cite{albert_boolean_2014, abou-jaoude_logical_2016, schwab_computational_2020, hemedan_computational_2022}. Here we will focus on the simplest discrete dynamic model, the Boolean model. This model uses two logic categories of node abundance. Any type of measurement that sorts the observations into two categories can be used to construct a Boolean model.  Boolean modeling is most appropriate if each variable has a bimodal distribution or if there exists a threshold abundance that must be exceeded to activate the node's interactions. The precise value of this threshold abundance does not need to be known because it does not enter the dynamic model. If research questions hinge on precise concentrations or high time-resolution (e.g., to determine dosing schedules), then a Boolean model might not be appropriate.

\section{Boolean modeling of the dynamics of biological networks}
In a Boolean model, each entity $i$ is characterized by a variable $X_i$ that can take one of two values: $1$ (``active'') or $0$ (``inactive''). Each variable $X_i$ updates its value according to the output of an update function $f_i:\{0,1\}^N\rightarrow \{0,1\}$ (also called a regulatory function) which equivalently maps each component of the system state $\boldsymbol{X}=(X_{i_0},\ldots,X_{i_{N-1}})$ to either $1$ or $0$. This map can be communicated via a so-called truth table, in which the output of every combination of function input values is listed, though this is often not efficient. Specific models may use different templates or representations for the regulatory functions, for example, threshold functions activate if a weighted sum of a node's regulator variables exceeds a specified value. Any Boolean function can be expressed algebraically using logical operators (e.g., ``not'', ``or'', ``and'' with symbolic representations $\neg$, $\vee$, and $\wedge$, respectively). Regulatory functions that do not depend on the regulated variable contain an implicit decay: if the regulatory function maps to $0$ the variable will be updated to the state $0$ even if currently its value is $1$.  

The specific timing of when to update each regulatory function also has an effect on the model. There are several schemes for determining the timing of variable updates. In the synchronous update scheme, all variables are evaluated at once in each discrete time step. This update scheme is realistic if the durations of synthesis and decay processes of each element are the same, or when external factors drive synchronization of system elements. For example, this may be the case for certain gene regulatory networks, as the time scale of gene transcription and mRNA degradation is similar, on the order of minutes \citep{alon2019} or in ecological networks where seasonal variations synchronize population growth. In contrast, asynchronous update schemes allow different nodes to update with different rates, which is necessary for networks that include multiple time scales. There are many ways to implement an asynchronous update. Some are deterministic, for example updating nodes according to a fixed order; others are stochastic, for example including a propensity for a variable to retain its current value. In the random order asynchronous update scheme, all nodes are updated once per round in a randomly selected order. Unless explicitly noted otherwise, we use the stochastic asynchronous update scheme, in which at each step a single variable is randomly chosen to update its value (each variable must have a non-zero update probability; these are often chosen to be uniform). Compared with other updating schemes, this scheme removes spurious oscillations that arise from unrealistic perfect synchrony, but otherwise preserves long-term behaviors \citep{saadatpour_attractor_2010}. Once the update functions are determined and an update scheme is selected, the Boolean system is fully specified.

Every trajectory of a Boolean model ultimately converges into a recurrent state or group of states, called an attractor. Attractors are divided into two types: point attractors (also called fixed points or steady states), which contain only one state, and complex attractors, which contain more than one state. In general, when using a stochastic updating scheme, the order in which the multiple states of a complex attractor are visited is not fixed. The attractors of models of biological systems correspond to stable biological behaviors, such as phenotypes in biomolecular models or stable communities in ecological networks. For example, the attractors of a molecular interaction network may correspond to cell types and cellular behaviors (such as the cell cycle).

The study of Boolean models of biological networks began with the work of Stuart Kauffmann and Ren\'{e} Thomas in the late 1960s. Ren\'{e} Thomas formulated Boolean and more general logical models of gene regulatory networks, e.g. the network underlying the development of bacteriophage\citep{thomas_boolean_1973}. Stuart Kauffmann introduced the study of ensembles of synthetic Boolean models to evaluate the expected properties of gene regulatory networks\citep{kauffman_metabolic_1969}; these ensembles have been used fruitfully ever since to obtain insight into the attractor repertoire of biological systems. The very first models (the so-called NK models) had a regular network structure in which each of $N$ nodes has a fixed number of regulators $K$, and each Boolean update function was constructed by selecting an outcome with a fixed bias toward 1, $p$, for each combination of values of the $K$ regulators. (Usually, this includes functions that have fully-redundant inputs, so the number of regulators can effectively be less than $K$.) Over time, improved understanding of the structure of biological networks has driven the use of more heterogeneous synthetic networks, for example networks that use a prescribed long-tailed degree distribution \citep{aldana_natural_2003}. In addition, careful study of the principles of biological regulation led to the narrowing of the pool of Boolean update functions \citep{harris_model_2002}. As there are no or very few examples in which a regulator has a positive effect in one context and a negative effect in another, the update functions should be locally monotonic functions \citep{raeymaekers_dynamics_2002}. There is increasing evidence that the inputs to the function form a hierarchy in terms of determining the output: a certain value of one of the inputs determines the output, otherwise the combination of certain values of two inputs determines a different value of the output, and so on~\citep{subbaroyan_minimum_2002}. Such functions are called nested canalizing functions \citep{jarrah_nested_2007}. In the following, we will focus on Boolean models of specific biological systems. For an excellent, recent review of the use of Boolean model ensembles we refer the reader to \cite{bornholdt_ensembles_2019}.

\section{How to build and validate a Boolean network model of a specific biological system} 
\label{model_features}
Like any model, a Boolean network model is only valid insofar as it captures the essential structure and behavior of the real system being modeled, and it is only useful insofar as it facilitates meaningful insight into the behavior of the system. A good model has both explanatory and predictive power. It identifies the root cause of certain outcomes, e.g., it identifies which elements play a key role in determining a certain attractor. It also answers ``what if?'' questions by identifying the outcome of situations that were not previously studied. In the context of Boolean networks, such explanations and predictions often come from the analysis of interventions, where the effect of modifying update functions and/or fixing nodes into the ON or OFF state (temporarily or permanently) is determined. A deciding factor in choosing the resolution or scope of the model is the expected value of the results of this perturbation analysis. Simply put, if a node is not included in the model then the model cannot predict what happens when the node (and/or its update function) is perturbed.

With this end goal in mind, the key features of a Boolean model of a specific system can be determined by considering the following questions:
\begin{enumerate}
    \item {\bf What should the nodes and edges represent?} While this is often self-evident from context, subtleties exist. For instance, in a protein-protein interaction network, a chain $A \rightarrow B \rightarrow C$ may indicate that protein $A$ has some influence on protein $B$, which in turn has some influence on protein $C$. If, however, protein $B$ has no other interactions in the network, it is often convenient to simplify the network by removing node $B$ and connecting $A$ to $C$ directly: $A \rightarrow C$. Thus, the edge from $A$ to $C$ implicitly incorporates protein $B$, making the distinction between a node and edge muddied. Moreover, directly connecting $A$ and $C$ can impact network dynamics, for instance by changing the convergence time into an attractor, or in rare cases even changing the attractors \citep{naldi_dynamically_2011, naldi_linear_2022}. In general, care must be taken to identify what nodes and edges represent, both for the sake of methodological clarity and when interpreting model behavior. 
    \item {\bf How are the nodes and edges identified?} Ideally, empirical data on the specific system is used for network construction. The data might be from direct observation (for instance, animal-plant interactions in ecological systems), experiments (for example, knocking out a gene and observing the effects on other gene products), or from literature review. Usually, the process of generating a network from the data is not straightforward. For instance, animal-plant interactions will occur with varying frequencies, so a modeler is required to either (a) choose a cutoff below which an edge is \textit{not} added to the network or (b) work in a modeling framework where edges carry weights. Additionally, perturbation experiments yield causal effects that may not correspond to direct interactions but to chains of such interactions; seemingly independent effects may in fact share mediators \citep{li_predicting_2006,maheshwari_inference_2022}. An increasing number of methods and tools exist to generate networks from large compilations of experiments  \citep{razzaq_computational_2018,munoz_griffin_2018}. For example, gene regulatory networks can be inferred from gene expression (transcriptomic) data using statistical or information-theoretic approaches, as reviewed by \cite{saint-antoine_network_2020}. Brain networks can be inferred from neuronal recordings based on statistical or model-based methods, as reviewed by \cite{magrans_connectivity_2018}. 
    If there isn't sufficient empirical information to construct a specific network model, synthetic models generated according to various algorithms can also bring insight. 
    \item {\bf How will system configurations be abstracted into a Boolean framework?} Dynamic quantities associated with a node are often implicit in the choice of nodes (e.g., concentrations for molecular species). In cases where the quantity is not inherently Boolean, the modeler must interpret the viability and meaning of a Boolean framework. The binarization of a continuous variable is usually done by imposing a threshold; below-threshold values are interpreted as the Boolean 0 and above-threshold values are interpreted as 1. In other cases, differential analysis (e.g. finding that a gene is expressed much more highly in a context compared to another context) is the basis of the binarization \citep{pandey_Boolean_2010}. 
    \item {\bf How will the update functions be determined?} The Boolean update function of each node includes the node's regulators (the starting points of edges incident on the node) as inputs, and should respect the positive or negative nature of the regulation if its sign is known. There usually are multiple possibilities consistent with the network, which are distinguished in the way the effects of multiple regulators are aggregated, e.g., whether two activators act independently or synergistically. Empirical information on the outcome of the function in cases of given input combinations can be used to find the best choice, or narrow the possibilities. This can be automated to some extent \citep{razzaq_computational_2018,munoz_griffin_2018,wooten_systems-level_2019}. In the absence of concrete information, observed general principles of biological regulatory functions can also be used to narrow the possibilities. As discussed in Sections 2 \& 3, many approaches exist to express the Boolean update functions, including logical statements (e.g. $f_A = B \wedge C$) that are most usefully expressed in disjunctive normal form, and threshold rules, where edges carry a numerical score and a node's state depends on the sum of incoming edge weights. The choice is highly context-dependent; logical models are appropriate where different combinations of inputs can lead to vastly different behavior, as is often the case in biomolecular networks, while threshold rules are appropriate when a node's behavior depends on the extent to which an interaction occurs and not directly on what other node caused the interaction (e.g., predation in food webs). Note that every threshold Boolean function can be expressed as a logical equation, but the reverse is not always true.    
    \item {\bf What is the most appropriate implementation of time?} The stochastic asynchronous update is appropriate in many cases, especially when the rates of specific interactions are unknown. Other choices may be appropriate if, for instance, interaction rates \textit{are} well known. Sometimes, the synchronous update may be preferred because it generally leads to faster convergence time and easier implementation of brute-force attractor identification, or because it enables the use of convenient mathematical tools. It can also be appropriate when external factors drive synchronization in the system, for example, when considering seasonal or annual factors in ecological models.
    \item {\bf How is the model to be validated?} The answer to this question depends on the available information about the system being modeled. Generally speaking, the static network (what are the nodes, and which nodes interact) should be validated first to make the success of the dynamic model (\textit{how} the nodes interact) more likely. The validation of the network entails general considerations (does the model network have a realistic structure, as measured by any number of standard or system-specific metrics?) and verification of specific edges. The validation of the dynamic model is done by verifying that it captures expected/known behavior, for example its attractors correspond to known phenotypes. If there is agreement for certain phenotypes but not for others, the discrepancy can sometimes be resolved by making a small change to the static network (e.g., assume a connection between two previously disconnected nodes). This yields new hypotheses and testable predictions; for example, a previously undiscovered inhibitory edge in a plant signaling network was predicted in this way and validated experimentally \citep{albert_new_2017,maheshwari_model_driven_2019}.
    \item {\bf What is the intended predictive power of the model?} That is, what is its intended use? This is highly context-dependent and, as mentioned above, feeds into the initial design of the model. In an ideal scenario, the model makes meaningful and testable predictions. For example, the model may indicate the phenotype repertoire of the system and predict how it changes in response to interventions. Empirical testing of the predictions provides additional validation or opportunities to refine and improve the model. Ideally, model predictions can be used to identify and prioritize therapeutic strategies that eliminate pathological phenotypes and drive the system toward desired phenotypes.
    
\end{enumerate}

There are many computational tools and analytical techniques developed for the analysis of Boolean networks. In subsequent sections, we will discuss several of these techniques in more detail. General purpose Boolean analysis software includes GinSim \citep{naldi_logical_2018}, CANA \citep{correia_cana_2018}, and pystablemotifs \citep{rozum_pystablemotifs_2022}. Many software tools are available in the CoLoMoTo repository of Boolean analysis Python libraries \citep{naldi_colomoto_2018}.

\section{Case study Boolean systems} 
In this section we discuss the formation of specific Boolean models under the framework outlined in the previous section. We outline considerations taken in the construction of these models, as well as the processes by which they were validated. Generally, the risk of over-parameterization is much smaller in Boolean models than for ODE or PDE models. Nevertheless, the accuracy and appropriateness of a Boolean model must be carefully evaluated by comparing with a compendium of experimental or observational results.

\subsection{A plant-pollinator model ensemble}
Plant-pollinator interactions are crucially important components of ecosystem stability in general and crop pollination in particular. Empirical plant-pollinator networks can be built, for example, through real-time observation of pollinator-plant visitation \citep{memmott_structure_1999} and/or trapping pollinators that visit particular species of plants \citep{russo_experimental_2019}. Precisely because they are empirical, such networks are immensely valuable when building and benchmarking simulated plant-pollinator networks. However, collecting high-quality data is very time-consuming and proper identification of interacting species requires a high level of expertise. A model capable of generating realistic ecological networks is therefore appealing, in part because of the ability to generate and statistically analyze many such networks (e.g., to determine the network properties that increase the likelihood that the community will collapse due to the loss of one or a few species). 

These considerations motivated the development of the bipartite Boolean model by \cite{campbell_network_2011}. The model is Boolean insofar as each species (i.e., node) \(X_i\) is assumed to be either present (1) or absent (0) in a community of interest at each time step \(t\). Plant-pollinator interactions are either mutually beneficial (pollinators feed from the plants and plants are pollinated by the pollinators) or beneficial for one species and wasteful for the other (pollinators spread the pollen but cannot feed from the nectar or pollinators can acquire nectar without spreading pollen). Mutually beneficial interactions are represented by a pair of positive edges connecting each species to the other, while one-sided interactions are represented by a positive edge to the species that benefits and a negative edge to the species for which the interaction is wasteful. Within the model, all species are assumed to persist in a so-called \textit{regional species pool} surrounding a community of interest. The composition of species in a community of interest changes in a synchronous update scheme such that species enter (from the regional species pool) or remain in the community at time \(t+1\) if their interactions with the species in the community at time \(t\) are favorable (i.e., if plants are able to be pollinated by species already in the community and if pollinators have sufficient food sources).

\subsubsection{Building the model}
The choice of representing the system with a network comprising species (i.e., plants and pollinators) as nodes and their interactions (i.e., pollination) as edges follows naturally from considering the system; the choice to represent the system as a \textit{bipartite} network follows from the fact that plant-plant and pollinator-pollinator interactions are largely irrelevant when considering pollination and pollinator feeding\citep{bascompte2003nested}. 

Once the choice to model the system using a network approach is chosen, the next decision to make is how to create and characterize the nodes and edges: how many of each type should exist in the regional species pool, which node pairs \(X_i\) and \(X_j\) should interact and which should not? A plant-to-polinator ratio of 1:3 was chosen to match empirical observations \citep{campbell_network_2011,campbell_whole_2022}. The model similarly draws from empirical values when generating edges: an exponentially cut-off power law fitted to empirical data is used to determine the degree distribution.

These interactions are ultimately used to determine, at a given time step, the Boolean state variable associated with each node, which corresponds to the relative abundance of that species in the community. We emphasize that if a variable is in the $0$ state, it does not necessarily imply that \textit{no} members of the corresponding species exist in the community. Rather, it implies that there are too few members of that species for their interactions to have an appreciable impact on the abundance of other species.

Indeed, the effect of each interaction is determined by a species-specific characteristic which is also assigned randomly according to empirical (skew normal) distributions: pollinators are assigned a \textit{corolla depth} and pollinators a \textit{proboscis length}. A close match (in practice, \( \leq 10\% \) difference) results in a mutually beneficial interaction; otherwise, the interaction is beneficial only for the species with the longer characteristic length. The rationale is that a pollinator must reach into the corolla to feed (which it can do with a sufficiently long proboscis), while a plant needs the pollinator to land on it for pollination to occur (which the pollinator will do unless the proboscis is so long the landing is not necessary). Note, however, that this framework does not take into account effects such as very small pollinators crawling or burrowing into the corolla.

\begin{figure}
    \centering
    \includegraphics[width=1\textwidth]{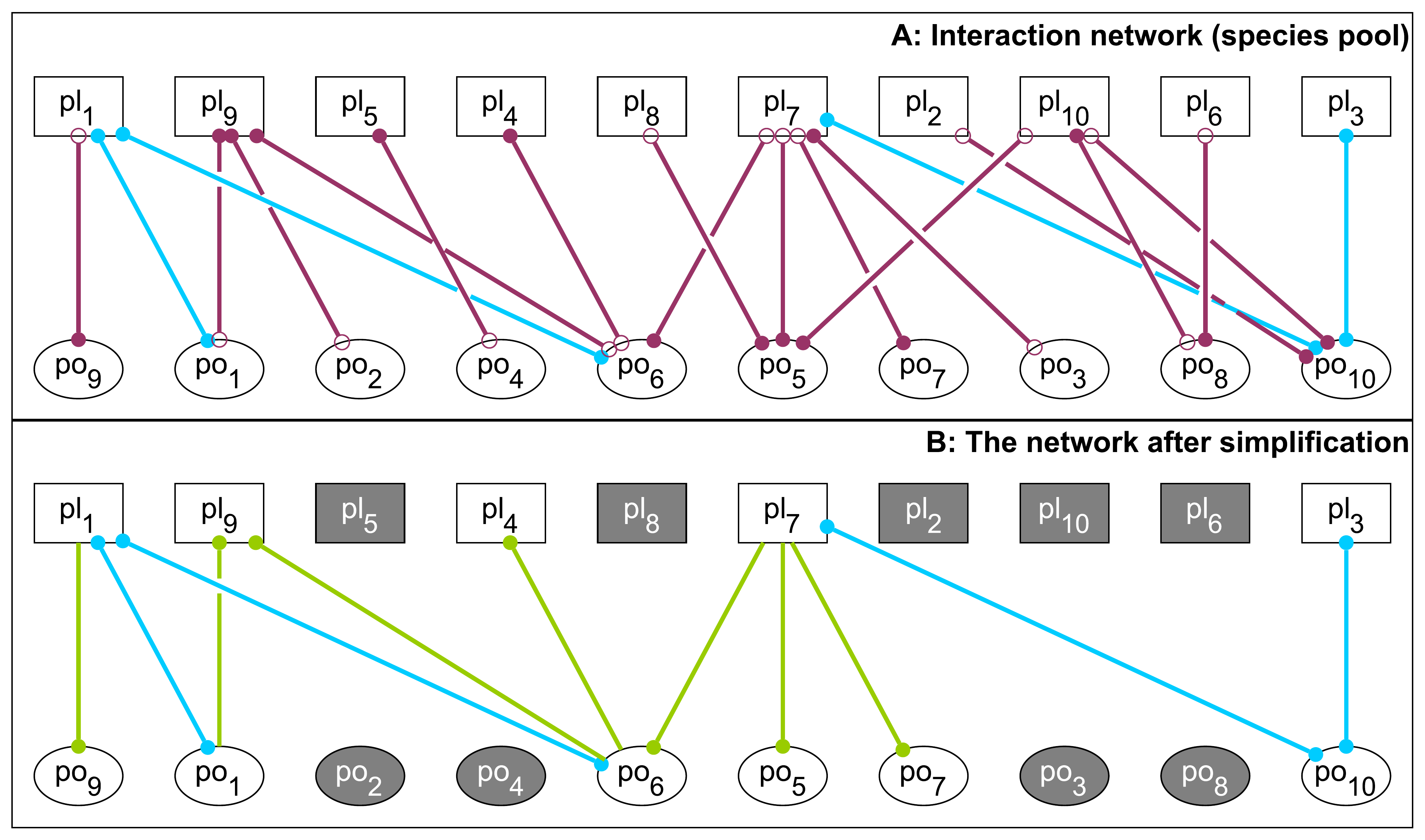}
    \caption{Example plant-pollinator network. This network consists of 10 plant species (labeled with ``pl'') and 10 pollinator species (labeled with ``po''). Each arrow ending in a solid circle denotes a beneficial effect on the target species, and each ending in a hollow circle represents a detrimental effect on the target species. Mutually beneficial interactions (the edges that are positive on both ends) are shown with blue; edges that are beneficial to one species but detrimental to the other are shown in purple. Panel A shows the species pool in which all species are included. Panel B demonstrates that some of these species (the ones with grey background) do not receive enough beneficial influence and hence cannot establish according to Equation ~\eqref{eq:bool_threshold}. According to Equation ~\eqref{eq:bool_threshold} the future state of the species that have at least one positive regulator and three or fewer negative regulators (e.g. node pl\textsubscript{7}) only depends on their positive regulator(s). This leads to the elimination of all negative edges in panel B, leaving unidirectional edges shown in green. }
    \label{fig:ppn_ex}
\end{figure}

With this framework in mind, the dynamic state variable of a node \(X_i\) at time \(t\) is given by

\begin{equation} \label{eq:bool_threshold}
X_i(t+1) = \left\{ \begin{tabular}{l l} 
1, & $\sum_j{E_{j i}X_j(t)} \geq$ 1 \\ 
0, & \text{otherwise} \end{tabular} \right., 
\end{equation}

where \(E_{ji} \in \{-1,0,4\}\) indicates a detrimental (-1), beneficial (+4), or nonexistent (0; i.e., no edge) interaction from node \(j\) to  node \(i\). In other words, a species $i$ persists in or colonizes the community of interest at time $t+1$ if the species in the community at time $t$ have a net positive impact on species $i$. The choice of numerical values is made in the absence of clear empirical data but reflects the fact a beneficial interaction (successful feeding or visitation by a viable pollinator) substantially outweighs the effect of negative interactions (e.g., wasted time and energy unsuccessfully attempting to feed from a plant). An example from the ensembles of \cite{campbell_network_2011} is provided in Figure~\ref{fig:ppn_ex}. This network consists of 10 plant and 10 pollinator nodes. The beneficial interactions are shown using arrows with solid black circles while the edges that represent detrimental interactions end in hollow circles.

The choice to implement a synchronous update scheme reflects a unified (and simplified) framework where species tend to emerge, for instance, at the beginning of a year/season, and the loss of some species may not be fully realized until the end of a year/season. While this framework ignores intra-seasonal dynamics (e.g., \cite{russo_supporting_2013}), it is perhaps appropriate given the prior modeling choice to represent species in a Boolean framework, which already flattens dynamic variations in population number. 

Analysis of Equation~\eqref{eq:bool_threshold} reveals several implications of ecological significance:
\begin{enumerate}
    \item A species that does not receive any beneficial influence from the species pool is not able to establish. An example of such species is node pl\textsubscript{6} in Figure~\ref{fig:ppn_ex}. Any other species whose sole positive influence is from such a transient species is also unable to persist (see node po\textsubscript{8} and pl\textsubscript{10}). The nodes that represent the species that are unable to persist are highlighted with grey in Figure~\ref{fig:ppn_ex}B. 
    
    \item If a species receives at least one positive influence and three or fewer negative influences from the community, its future state only depends on the current state of its positive regulator(s). As a result, the negative edges become irrelevant and the network representing the community simplifies further. For instance, all the negative edges incoming to node pl\textsubscript{7} have been removed in the simplified network shown in Figure~\ref{fig:ppn_ex}B.
\end{enumerate}
These observations lead to a much simpler network which shows that only a fraction of the original species pool has the opportunity to establish and persist in the community. The resultant simplified network is shown in Figure~\ref{fig:ppn_ex}B, in which the isolated nodes highlighted with grey represent the species that cannot establish, and the superfluous negative edges are removed. These simplifications reduce the network from 20 nodes and 38 edges to 11 nodes and 15 edges without changing the dynamics.

\subsubsection{Model validation and applications}
The attractors of the model correspond to stable communities: the nodes whose state variables in the attractor are not fixed in the state 0 are part of the community. The stable community contains a subset of the original bipartite network. The properties of these network subsets can be measured \citep{campbell_network_2011}; similarities to empirical networks lend confidence to the predictive power of the model. For example, many real plant-pollinator communities contain generalists (species that interact with many other species) and specialists (species that interact with few other species) in a so-called \textbf{nested} structure where specialists primarily interact with generalists \citep{bascompte2003nested}. While high nestedness was commonly held to promote stability, the model suggested that an over-reliance on nestedness made the community particularly susceptible to collapse if certain generalist species were removed from the community \citep{campbell2012topology}. The implication is that high nestedness can negatively impact stability, as also argued by others \citep{allesina_stability_2012, staniczenko_ghost_2013}.

The model recapitulates field observations in the case of perturbation of plant-pollinator communities. For example, a meta-analysis by~\cite{vila2011ecological} revealed that the invasion of alien plants into an established community caused a considerable decrease in native plant and animal species abundance. The same behavior is observed in this model: invasive plant species that attract pollinators using mimicry while not providing food for them add negative interactions to the system and prevent the formation of stable groups of species, resulting in decreased species richness and abundance in the community. The model also captures the extinction cascade following the extinction of generalist species \citep{biella2020empirical}. 

The model predicts how communities re-stabilize after the introduction or loss of species. Some notable findings include:
\begin{enumerate}
    \item The model captures the ability of some species to re-invade the community from the regional species pool if they are only removed from the community of interest, and predicts that the extent to which the targeted species behaves as a generalist in the host community is an important predictor of the long-term effect of its removal from the local community \citep{labar2013global}.
    \item In the context of restoring communities that have suffered a partial collapse due to species loss, the model predicts that the introduction of multiple generalist species can help to restore biodiversity, though the resulting community may be very different from the original community \citep{labar2014restoration}. 
    \item The model suggests that the physical characteristics of an invasive species (e.g., proboscis length for an invasive pollinator) can drive qualitatively different ecological responses depending on whether they are \textit{typical} vs. \textit{atypical} of native species \citep{campbell2015plant}. It also predicts that groups of species that simultaneously invade can have drastically different impacts on the host communities if the invaders are randomly selected compared to when they constitute a separate stable community \citep{campbell2022whole}. 
    \item The model predicts the magnitude of cascading extinctions and significant damages in plant-pollinator communities. It also suggests effective mitigation measures to protect the environment against the threat of species extinction and identifies fruitful community restoration strategies~\citep{fatemi2023predicting}.
\end{enumerate}

\subsection{Boolean models of signal transduction networks} 

Our next case studies will be two Boolean models of signal transduction networks that determine cell phenotypes. Both models were constructed via the integration of experimental evidence available in the literature. The first step in this construction process is to decide what nodes to include in the network. Since both networks focus on an identifiable phenotype, the known markers or determinants of this phenotype are included as nodes. The known signals (external molecules) that induce or affect the phenotype are also included and indicate natural inputs (source nodes) to the network. Each of the two networks integrates the interactions and regulatory relationships among their nodes as directed and signed edges. Both networks include an abstract sink node in the network as a marker of the phenotype (although the dynamic model will characterize the phenotype in more detail, including the state of other nodes). Each network is checked for consistency with known information. For example, every node that is known to affect the phenotype in a positive or negative way should have a positive or negative path, respectively, to the phenotype marker node.  

The next step is to determine the regulatory functions; this is also informed by experimental results that identify interactions and causal effects. When there are multiple regulators for a node, the modeler selects the function that best represents the existing knowledge about the regulatory action. The OR function is used if the node can be activated by any of its regulators. The AND function is used if the node needs all of its regulators to be activated. If a Boolean regulatory function involving several regulators cannot be fully determined, one needs to take an exploratory approach to select the function that can successfully reproduce the existing experimental results (both at the node and at the whole network level). 

Both models underwent an extensive verification process in which model dynamics were compared to experimental results and known biological information. In cases when one of the models did not recapitulate an experimental observation (e.g. the reduction of the level of a protein when a gene was knocked out) the modelers scrutinized the literature further and identified possible changes to the regulatory functions or to the interaction network that were consistent with the literature and helped align the model with the experiments. The models validated by this process were then used to make novel predictions of the cell phenotype in cases that were not studied experimentally. As both of these models concern pathological phenotypes, the predictions of most interest are therapeutic interventions that would eliminate the pathological phenotype.

\subsubsection{T cell large granular lymphocyte leukemia model} \label{TLGL}

\begin{figure}
    \centering
    \includegraphics[width=1\textwidth]{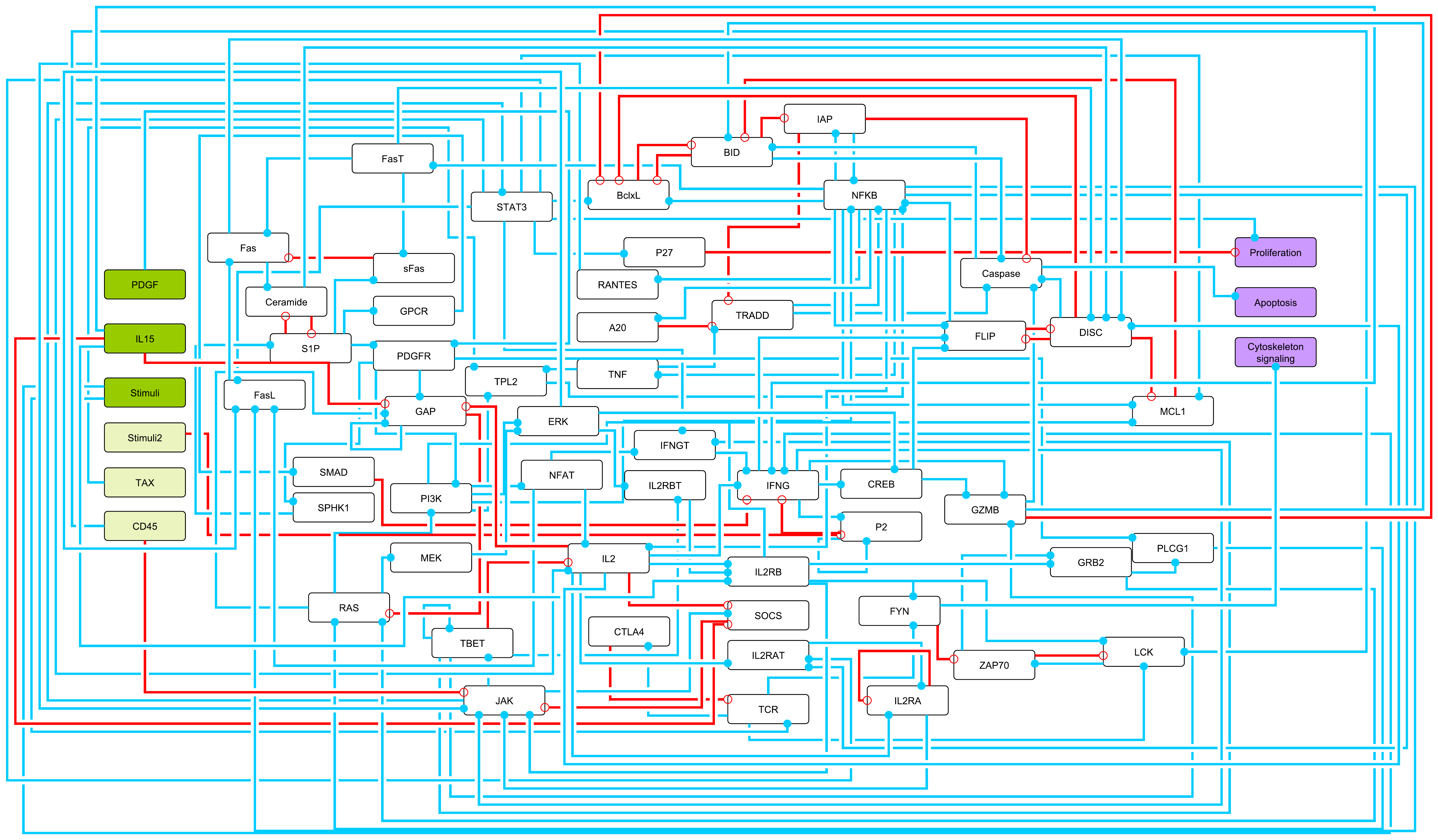}
    \caption{Interaction graph of the T-LGL model~\citep{zhang_network_2008}. The source nodes that represent external signals are shown in green, the source nodes that represent a cellular context are light green, and the three sink nodes that represent phenotype markers are purple. The Apoptosis node implicitly inhibits all other nodes in the network because it indicates cell death.}
    \label{fig:TLGL}
\end{figure}

T cell large granular lymphocyte (T-LGL) leukemia is a rare blood cancer. The T-LGL model of Zhang and collaborators \citep{zhang_network_2008} aims to capture the origin of this disease, which is the survival and proliferation of cytotoxic T cells. Cytotoxic T cells are generated to fight a microbial infection by eliminating infected cells, and after the infection is cleared they usually undergo the process of activation-induced cell death (apoptosis). Through an extensive literature search, Zhang et al. synthesized the nodes, interactions, and pathways involved in activation-induced cell death, in cell proliferation, as well as the pathways that were at the time known to be different in T-LGL cells compared to normal cytotoxic T cells (see Figure \ref{fig:TLGL}). 
The 58 nodes of the network include proteins, mRNAs, and small molecules, as well as the phenotype marker nodes “Cytoskeleton signaling”, “Proliferation” and “Apoptosis”. The network includes three signals: Stimuli, which represents the initial stimulation of the cytotoxic T cell that yields its activation, the immune signaling molecule interleukin-15 (IL15), and platelet-derived growth factor (PDGF). These latter two signals were found to have a high abundance in the blood of T-LGL patients. The 123 edges represent transcription and its regulation, protein-protein interactions, and chemical reactions. 

The T-LGL Boolean model of cytotoxic T cell apoptosis incorporated literature-derived regulatory functions. As the node "Apoptosis" represents the death of the cell, Zhang et al. assumed that Apoptosis inhibits every other node, thus its activation leads to a state in which Apoptosis is on and every other node is off. These inhibitory edges beginning at the Apoptosis node were effectively made implicit in follow-up studies~\citep{zanudo_cell_2015}, which removed these edges from the model but considered every state in which the node Apoptosis is ON as equivalent apoptotic states. Following this convention, we do not depict these edges in Figure \ref{fig:TLGL}. The model was used to simulate the system's trajectories starting from an initial state that describes a recently stimulated T cell. As there was very little information on the timing of events, the simulations used stochastic timing, specifically random order update, and the results of hundreds of simulations were summarized into a frequency of node activation. The single restriction on the update order was that the phenotype node "Apoptosis" was updated first, thus the round of update that started with Apoptosis turning on ended by convergence into a point attractor in which every other node was off. The model has an additional attractor, namely a point attractor in which Apoptosis is off; convergence into this attractor signifies the survival of the cell, which is an abnormal behavior in the modeled context. The model reproduces the experimentally observed survival of a subset of the initial stimulated cells and the known markers of this process, for example the activation of JAK in every surviving cell. The model predicts that a small subset of the known dysregulations (abnormal node states in T-LGL) is sufficient to cause all the others, thus preventative efforts should focus on this subset. The model predicts the T-LGL state of 12 additional nodes that were not assayed experimentally. The model also predicts several key nodes whose state change can ensure that the entire cell population undergoes apoptosis; these key nodes are potential therapeutic targets for T-LGL leukemia. Several of these predictions, such as the apoptosis-inducing effects of inhibiting PDGF, NF$\kappa$B, or S1P, have been verified experimentally~\citep{zhang_network_2008,saadatpour_dynamical_2011}.

\subsubsection{Epithelial to mesenchymal transition model} \label{EMT}

Epithelial cells, examples of which include skin cells and cells in the lining of internal organs, form a sheet in which each cell is bound to the neighboring cells. During the process of epithelial-to-mesenchymal transition (EMT) epithelial cells lose their adhesive properties and transform into mesenchymal cells, which are separate and free to move. The EMT process is important in embryonic development and in wound healing, both of which involve significant tissue-level reorganization. This process can also be hijacked in cancer and forms the basis of cancer cells' ability to leave the primary tumor and initiate metastasis. The loss of the cell-cell adhesion protein E-cadherin (encoded by the gene \textit{CDH1}) is considered the hallmark of EMT. The transcription of this protein is downregulated by seven transcription factors. A variety of external molecules, mainly growth factors, can act as signals to this process. 

\begin{figure}
    \centering
    \includegraphics[width=1\textwidth]{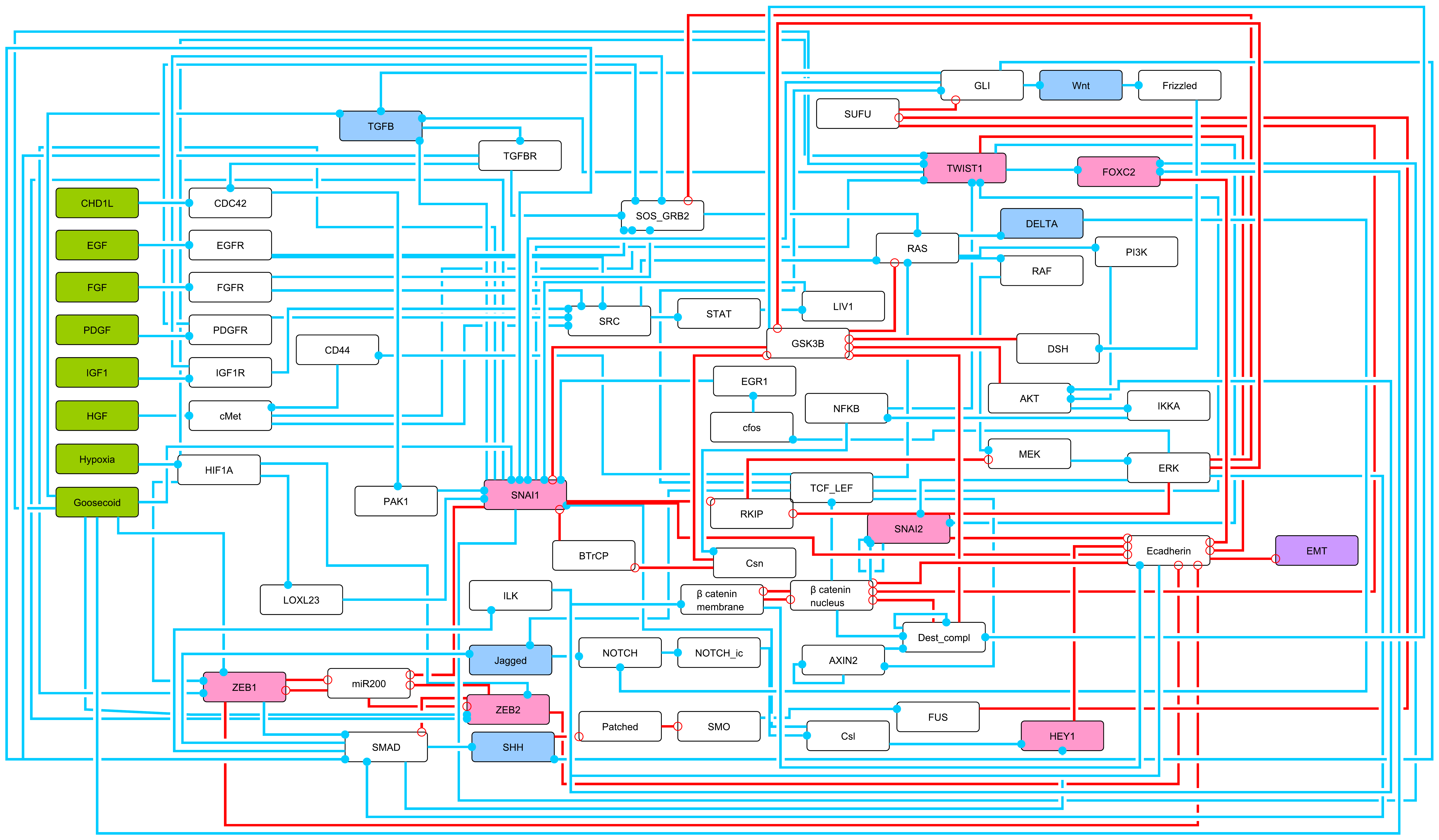}
    \caption{Interaction graph of the EMT model. The external signals that are source nodes of the network are shown in green, the molecules that can act as signals and are also produced by the cell are shown in blue, and the transcription factors that downregulate the transcription of E-cadherin (the hallmark of EMT) are shown in pink. }
    \label{fig:EMT}
\end{figure}

The EMT signaling pathway encompasses the chains of interactions and reactions that start from these signal molecules and end in one or multiple markers of the EMT transition. The transcriptional downregulation of E-cadherin is one such marker; another frequently used marker is the transcriptional upregulation of the protein vimentin. Steven Steinway and collaborators synthesized the experimental evidence into a 69-node, 134 edge network (see Figure \ref{fig:EMT}) that contains 13 signal nodes and ends in the transcriptional downregulation of E-cadherin as a marker of the EMT transition~\citep{steinway_network_2014}. Five of the signals can also be produced by cells that undergo the EMT process, forming a positive feedback loop; the other eight are treated as source nodes that encode an external context. Steinway et al. used the experimental literature to determine the Boolean regulatory functions of each node. The main goal of Steinway et al. was to model EMT in the context of hepatocellular carcinoma (liver cancer). As the growth factor TGF$\beta$ is known to be aberrantly produced by liver cancer cells, the model assumes that TGF$\beta$ is in a sustained ON state. Signals that are source nodes are assumed to be in a sustained OFF state, and signals that participate in positive feedback loops are initiated in the OFF state (but have the possibility of turning on during the process). The model uses a stochastic asynchronous update scheme in which nodes are grouped into two categories: nodes that are regulated by signal transduction (protein-level) events are expected to change state faster, and thus are updated with a higher probability, than nodes that are regulated transcriptionally. To simulate the EMT process, Steinway et al. started an ensemble of model copies from an initial condition representative of a prototypical epithelial cell, then turned TGF$\beta$ on. All the replicate simulations followed a similar trajectory and converged into a point attractor that recapitulates the known features of mesenchymal cells. The model recapitulated known dysregulations observed during EMT of cancer cells. It also predicted the activation of two additional pathways (Wnt and SHH), which were previously thought to be independent of TGF$\beta$; this prediction was validated experimentally \citep{steinway_network_2014}.  

\section{How to analyze a Boolean model: state transition graphs, attractors, and trap sets}

Now that we have established the general principles of Boolean model construction and illustrated them with three examples, we are ready for more detail on how to analyze and obtain insights from a Boolean model. In this section we will describe traditional approaches to comprehensively map the state space of Boolean models. The next sections will describe alternative, network-based approaches that - as we will illustrate - are equally insightful and more computationally efficient.

\begin{figure}
    \centering
    \includegraphics[width=1\textwidth]{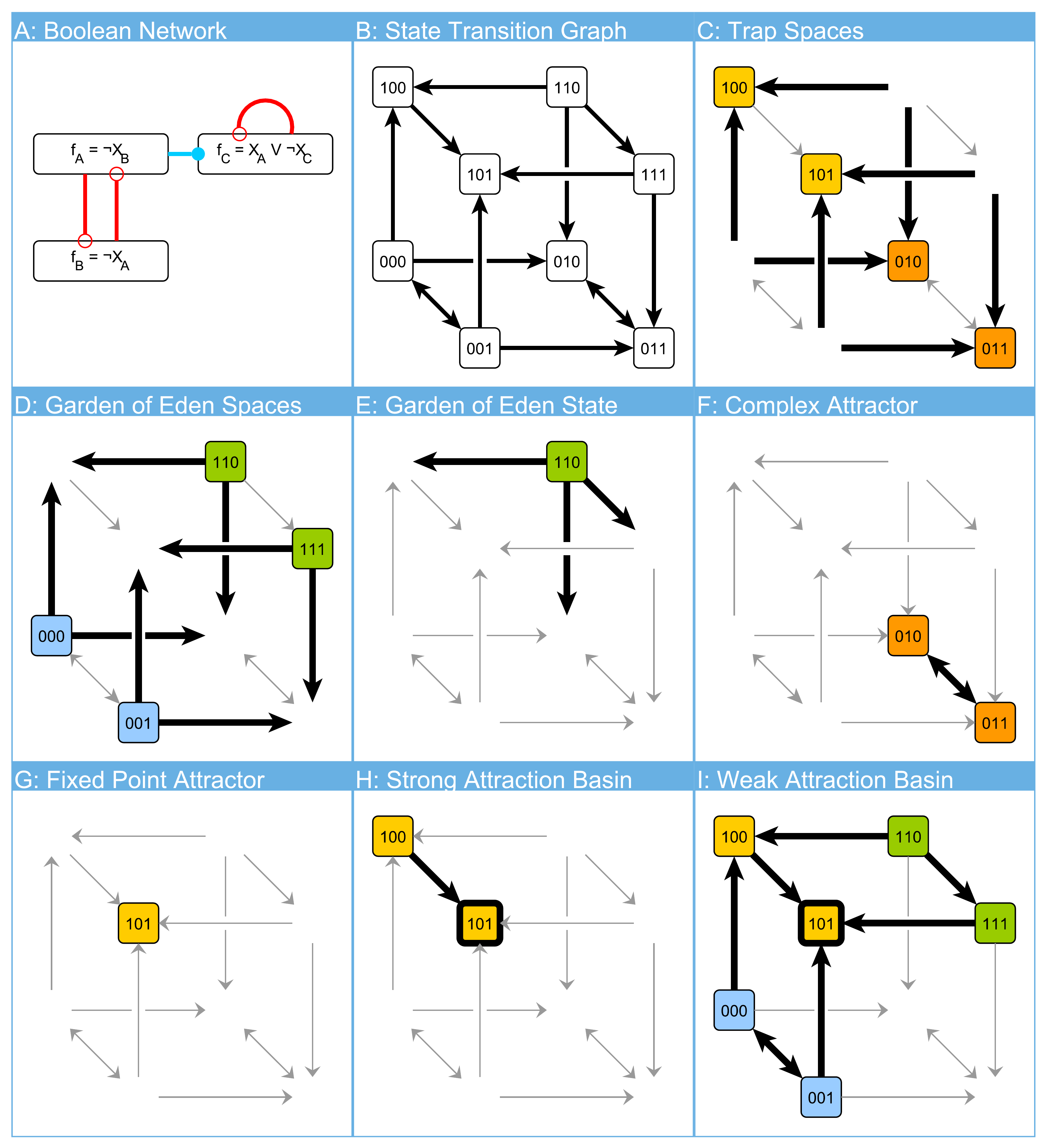}
    \caption{State transition graph example with special subsets highlighted. In panel A, the Boolean network and its regulatory functions are depicted. In panel B, the corresponding STG for the asynchronous update scheme is shown. Each node represents a system state by specifying the states of the three Boolean variables $X_A$, $X_B$, and $X_C$, in order. Edges represent allowed transitions (transitions from a state to itself are not drawn explicitly, following standard conventions). Panel C highlights two trap spaces, $10\star$ in yellow, and $01\star$ in orange. Neither pair of states has transitions that leave the pair; incoming transitions are highlighted. Panel D highlights two Garden of Eden spaces, $11\star$ in green, and $00\star$ in blue. Neither pair of states has transitions that enter the pair; outgoing transitions are highlighted. Panel F shows an attractor of the system: a minimal set of states from which there is no escape. Another attractor, a fixed point attractor, is shown in panel G. Its strong basin of attraction, or the set of states that can reach no other attractors, is shown in panel H. In panel I, the weak basin of attraction for the fixed point, i.e., the set of states that can reach the attractor, is highlighted.}
    \label{fig:STG}
\end{figure}

The Boolean update functions and the updating scheme together give rise to a \emph{state transition graph} (STG) with $2^N$ nodes representing all possible system states. Directed edges from one node (system state) to another indicate that the parent state can be updated in one time-step to attain the child state (see Figure \ref{fig:STG}). Under an asynchronous update scheme in which a single randomly selected variable is updated at each time-step, each node of the STG has between 0 (in the case of a steady state) and $N$ (in the case when each node changes state when updated) outgoing edges. In a deterministic synchronous update scheme, each node has only one outgoing edge, because all the nodes are updated at the same time. The attractors of a Boolean system are the terminal strongly connected components of the STG (i.e., they have no edges that exit the component). These are divided into two types: point attractors (also called fixed points or steady states), which contain only one state, and complex attractors, which contain more than one state. Sometimes, it is useful to further subdivide the class of complex attractors into two subclasses: limit cycles, in which each state has only one possible successor, and disordered attractors, in which some of the attractor states have multiple possible successors. The in-component of each attractor, referred to as the basin of attraction of that attractor, indicates the states for which there exists at least one trajectory that converges to that attractor. If the update scheme is nondeterministic, then multiple attractors can be reachable from the same state. In these cases, it is important to distinguish an attractor's weak basin of attraction from its strong basin of attraction. An attractor's weak basin of attraction is the set of states from which the attractor is reachable, while its strong basin of attraction is the set of states from which the attractor \emph{and no other attractor} is reachable. Importantly, the topology of the STG is not affected by biasing some nodes to update more frequently than others. Therefore, the attractor repertoire does not depend on the precise probabilities that individual nodes are selected for update in the stochastic asynchronous update, as long as the probabilities remain nonzero.

The size of the state transition graph grows exponentially with the number of variables, so directly analyzing the STG is not always sensible. For medium to large Boolean networks, it is often not feasible to construct the full STG; a network with 50 nodes requires thousands of terabytes of storage to produce. For very large networks, constructing the STG is not only infeasible, it is impossible. The STG of a 265-variable network has approximately as many nodes as there are atoms in the observable universe. Even when a Boolean network is relatively small and its STG can be constructed, it is often difficult to analyze the STG and to extract biological meaning from it. Insofar as such endeavors are fruitful, they are usually equivalent (or practically equivalent) to simulating many copies of the Boolean network using different initial conditions and update orders. Various tools exist for this task of simulating Boolean dynamics and sampling the STG \citep{albert_boolean_2008,correia_cana_2018,naldi_logical_2018,park_robustness_2023}.

Ultimately, the STG and the dynamics of a Boolean network are equivalent, and therefore the STG is the primary mathematical object of interest we wish to characterize. However, its unwieldy nature makes indirect approaches considerably more desirable than brute-force STG construction. In particular, a common goal is to identify key subgraphs of the STG without explicitly constructing them. As mentioned earlier in this section, the attractors are one class of subgraph that is of interest. Trap sets are another type of subgraph that is dynamically important. A trap set is a collection of nodes in the STG from which there is no escape; all paths that begin in a trap set remain in the trap set. In the special case when the trap set forms a space, i.e., when it can be fully characterized by specifying a set of constant variables, the trap set is called a trap space (see Panel C of Figure \ref{fig:STG}). A trap space (or, more generally, any subspace) is often specified using a "star" notation, in which constant variables are specified and free variables are labeled with a $\star$ symbol. For example, $10\star\star$ corresponds to the set of system states $\{1000,1001,1010,1011\}$. A trap set or trap space describes a possible dynamical commitment to a particular range of fates. A related concept is that of a Garden of Eden set, which is a set that cannot be reached from any starting point outside the set (see Panel D of Figure \ref{fig:STG}). These sets describe configurations that are not globally stable, as once a system in a state belonging to a Garden of Eden set is sufficiently perturbed, it can never return to that Garden of Eden set. Notably, if an attractor has any state inside a Garden of Eden set, that attractor must be entirely contained in that Garden of Eden set.

Various computational tools exist for identifying trap spaces \citep{klarner_pyboolnet_2016,petre_minimal_2022} and Garden of Eden Spaces \citep{rozum_pystablemotifs_2022}, and identifying attractors \citep{rozum_pystablemotifs_2022,benes_aeonpy_2022,trinh_computing_2022}. These problems contain the problem of identifying fixed points as a special case. Fixed points can be viewed as solutions to the SAT problem, which is NP-Hard. Thus, all these algorithms have worse-than-polynomial worst-case scaling. Reasonable run-times are obtained by the algorithms because they perform their exhaustive searches of the solution space in a manner that exploits the expected sparsity of the network to arrive at solutions faster than the average case for general networks. The precise average scaling for each algorithm will depend on the networks considered.

\section{The parity-expanded network}
In this section, we describe an approach to encoding regulatory logic in the structure of a so-called parity-expanded network; the aim is to identify dynamically important combinations of variable states without relying on computationally expensive STG construction or sampling. We will show that by analyzing the parity-expanded network, one can identify a hierarchy of trap spaces and construct the so-called succession diagram, which summarizes key features of the STG and can be used to identify interventions that drive the system to a target attractor.

The parity-expanded network, also called the logical or prime-implicant hypergraph, was introduced as an auxiliary network constructed from the Boolean update functions \citep{albert_topology_2003,zanudo_effective_2013}. The parity-expanded network is a hypergraph whose nodes represent Boolean literals (e.g., $X_i$ and $\neg X_i$). Hypergraphs connect nodes using hyperedges, which are generalized edges that connect one set of nodes to another, rather than only connecting one node to one node. The hyperedges of the parity-expanded network represent prime implicants of the update functions and their negation. For instance, a hyperedge incident on the node representing $\neg X_i$ represents an irreducible set of regulator states that result in the output $f_i(X)=0$ in the update function for $X_i$. Here we follow \cite{rozum_parity_2021} and introduce the expanded network using parity-related concepts and highlight its role as an invariant of the parity transformation.

The parity transformation acts on a Boolean system by the change of variables $X_i \mapsto \neg X_i$ for all variables $X_i$, mapping the original system with update functions $f_i$ to the system governed by update functions $\neg f_i$. This mapping induces further transformations on any structure derived from the update functions, and so, in a slight abuse of notation, we say the parity transformation acts on all these structures. For example, the parity transformation relabels the nodes of the state transition graph so that all 1s become 0s and vice versa. Viewing these node labels as spatial coordinates (so states lie on the vertices of a unit hypercube), the parity transformation is the spatial inversion of this hypercube through its center.

A parity-expanded network $G$ is a dynamically endowed hypergraph. Each node $I$ of the parity-expanded network, called a virtual node, is an ordered pair $I=(n(I),s(I))$ consisting of a system entity, in this context denoted $n(I)$, and a value $s(I)$, which is either the constant $1$ or the constant $0$. Since in the usual situation the interaction network is constructed first, in practice $n(I)$ is denoted by the name of the relevant entity. For example, one of the virtual nodes in the parity-expanded network for the T-LGL Boolean network is (Apoptosis, 1), which represents the activation of ``Apoptosis''. There are two virtual nodes associated with each system entity $i$, namely $(i,1)$ and $(i,0)$. We call this pair of virtual nodes contradictory to refer to the fact that a system entity cannot be in two states at the same time. A set of virtual nodes that does not contain any contradictory pairs is called consistent. We define the parity operator on a set of virtual nodes by $\neg S=\{(i,1-s): (i,s)\in S\}$. 

The parity-expanded network is a dynamical system in its own right: each virtual node $I$ is endowed with a Boolean variable $\sigma_I$, whose time evolution is governed by an update function $F_I$ which is obtained from $f_i$ for $I=(i,1)$ or from $\neg f_i$ for $I=(i,0)$. This function defines a $2N$-dimensional dynamics that can be restricted to the N degrees of freedom of the underlying Boolean system if we require that no pair of contradictory virtual nodes is ever simultaneously active.  In the context of this restriction, we may think of $\sigma_I$ as the indicator function for the subspace defined by $I$, i.e., we define  $\sigma_{(i,0)} (X)=\neg X_i$, and $\sigma_{(i,1)} (X)=X_i$. We view the expanded network as having two layers: an “original update” layer consisting of virtual nodes of the form $(i,1)$, and a “parity update” layer consisting of virtual nodes of the form $(i,0)$. The main advantage of this perspective is that it allows us to treat the negated and non-negated versions of variables and functions simultaneously.

\begin{figure}
    \centering
    \includegraphics[width=1\textwidth]{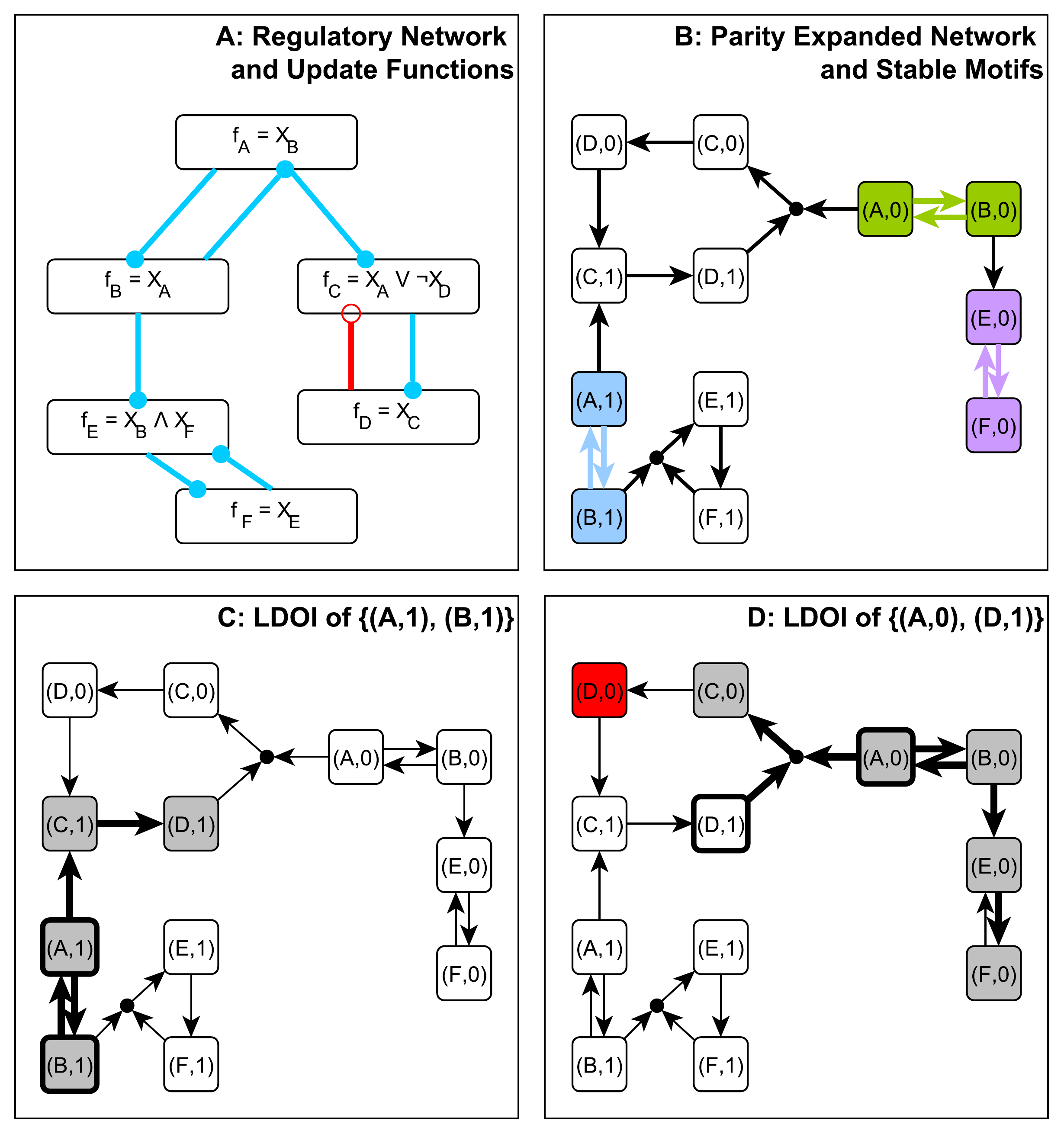}
    \caption{The parity-expanded network, stable motifs, and logical domain of influence. A Boolean network is depicted with its regulatory functions in panel A. The corresponding parity-expanded network is given in panel B, and the stable motifs are highlighted in color. In panel C, the logical domain of influence (LDOI) of one of the stable motifs is shown, highlighted in grey. The seed set is highlighted in bold outline. Notably, the LDOI of the stable motif includes itself, which gives rise to self-sustaining behavior in the system when A and B both activate. In panel D, the LDOI of a self-negating set is shown, with the contradiction boundary highlighted in red. When A is off and D is on, C will inactivate, which leads to the inactivation of D; thus A=0, D=1 is not a self-sustaining configuration. Nevertheless, panel D shows that by exogenously sustaining this configuration, B, C, E, and F will all eventually inactivate.}
    \label{fig:PEN}
\end{figure}

The dynamics of the $\sigma_I$ activity variables, and thus of the original Boolean network, are encoded in the connectivity of the parity-expanded network. A hyperedge connects a set of parent virtual nodes $S=\{I_0,I_1,\ldots,I_k \}$ to a target virtual node $J$ if $\bigwedge_{I\in S} \sigma_I$ is a prime implicant of the update function for $\sigma_J$, i.e., of $F_J$. For example, a hyperedge from the set $\{(1,0),(2,1),(3,0)\}$ to the virtual node $(4,0)$ indicates that $\neg X_1 \wedge X_2 \wedge \neg X_3$ is a prime implicant of the negated update function $\neg f_4$, meaning that $X_1=X_3=0$ and $X_2=1$ is a sufficient and irreducible condition for the output of $f_4$ to be $0$. Pictorially, we represent hyperedges with more than one parent using intermediary “composite nodes”, which correspond to “and” gates. For an example of a Boolean system and its parity-expanded network see Figure \ref{fig:PEN}. Hyperedges between and within parity layers encode important features of the dynamics. For example, negative influence manifests as inter-layer hyperedges. Thus, if a Boolean system’s interaction graph lacks negative feedback loops and has no paths of opposite sign between any two nodes, then there is a change of variables that disconnects the parity layers from one another \citep{rozum_parity_2021}. Each hyperedge represents logical sufficiency, i.e., if the virtual node source(s) of the hyperedge is(are) active, the target of the hyperedge will eventually become active. 

Because the parity-expanded network is constructed by computing the Blake canonical form of each update function and its negation, all prime implicants must be identified. The number of prime implicants, in the worst case, scales exponentially in the number of inputs. In general, computing the parity-expanded network is NP-Hard. Nevertheless, it is a tractable problem in sparse networks, which are common in biomolecular applications.

\subsection{Stable motifs}

Arguably the most dynamically important of the parity-expanded network’s topological structures are its stable modules and stable motifs \citep{zanudo_effective_2013,rozum_parity_2021}, which are subgraphs of the parity-expanded network that correspond to specific states of generalized positive feedback loops in the interaction graph of the Boolean network. These feedback loop states are self-sustaining and describe trap spaces in the dynamics. As described in Section 6 and illustrated in Figure \ref{fig:STG}C, trap spaces are regions of the state space characterized by a set of fixed variable values that, once attained by a trajectory, confine the trajectory to that region for all subsequent time steps. A stable module $M$ is a non-empty sub-hypergraph of the parity-expanded network such that every virtual node in $M$ has an incoming hyperedge included in $M$ (i.e., $M$ is sourceless) and no pair of virtual nodes in $M$ are contradictory (i.e., $M$ is consistent). A stable motif is a stable module that does not contain any smaller stable module; note that this implies that a stable motif is strongly connected. Because stable modules (and thus also stable motifs) are sourceless, every virtual node in a stable module $M$ can be maintained in its active state by other virtual nodes in $M$; because $M$ is consistent, this does not give rise to contradictory activation of virtual nodes. Together, these properties imply that the activity of $M$ is self-sustaining. Once a stable module is activated, it cannot be inactivated except via direct override of its virtual node activities by direct external controls (as opposed to inactivation via the effects of upstream pathways). In other words, a stable module describes a control-robust trap space in which the values of certain variables are stationary \citep{zanudo_effective_2013,rozum_parity_2021}.

Interpreting the meaning of the stable modules or stable motifs of a Boolean system depends on the context of the model. For example, stable motifs of plant-pollinator networks represent survival units. There are two main types of survival units. The first type is a group of species that, due to their mutualistic interactions, can sustain the above-threshold abundance of all species in the group; this is corresponds to a stable motif whose virtual nodes represent ON states of variables. The other type is a group of species that are present in the species pool but are absent from the community. Due to their interdependence, none of these species is able to join the community individually. This type of survival unit corresponds to a stable motif whose virtual nodes represent OFF states of variables. In principle, a third type of stable motif is possible: one in which some species are present, and some are absent, representing that for one species to infiltrate the local community, another must be absent. In the plant-pollinator networks considered here, such occurrences are extremely rare, as is expected in mutualistic networks \citep{campbell_network_2011,campbell_whole_2022}.   

\begin{figure}
    \centering
    \includegraphics[width=1\textwidth]{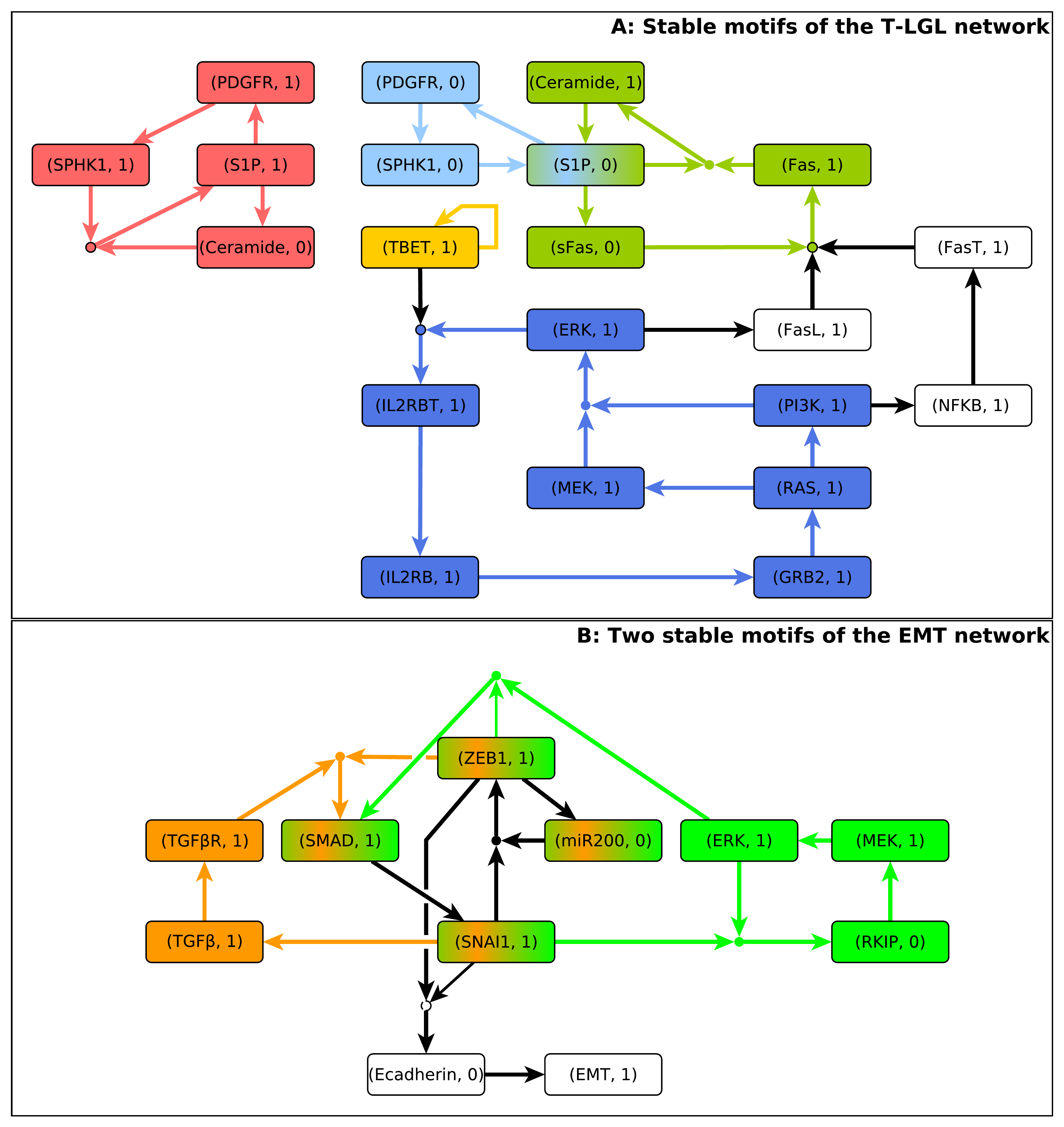}
    \caption{Illustration of stable motifs in the T-LGL and EMT models. Each motif is highlighted by the background color of the symbols and the color of the edges; black edges mean participation in multiple motifs. A: stable motifs and conditionally stable motifs of the T-LGL leukemia model. The dark blue and green motifs are conditionally stable. One way of identifying their direct conditions is by noting the external inputs to composite nodes of the motif. The indirect conditions are sets of virtual nodes whose LDOI contains the direct conditions. B: two of the eight stable motifs associated with epithelial to mesenchymal transition. In the full expanded network, the virtual nodes (SNAI1, 1) and (ZEB1, 1), as well as six additional virtual nodes, are incident on a composite node that targets (Ecadherin, 0). In this figure, the composite node is represented by an empty, dotted circle to indicate that they are not the only virtual nodes incident on this composite node. }
    \label{fig:TLGL&EMT_motifs}
\end{figure}

The T-LGL leukemia model, applied in the context in which antigen stimulation (Stimuli) and the cytokine IL15 are present and the growth factor PDGF is absent, has three stable motifs \citep{zanudo_cell_2015}. One corresponds to the OFF state of the receptor protein PDGFR, the enzyme SPHK1 and the product of the reaction catalyzed by SPHK1, S1P; the virtual nodes of this motif are marked by light blue background in panel A of Figure \ref{fig:TLGL&EMT_motifs}. Another stable motif combines the ON state of these three nodes with the OFF state of the protein Ceramide (red background in Figure \ref{fig:TLGL&EMT_motifs}A). Each of these stable motifs has a chance of activating in mutually exclusive trajectories of the system. We note that if the growth factor PDGF was present, it would activate PDGFR, meaning that the first stable motif would never activate. A third stable motif expresses the activity of the transcription factor TBET, which activates the transcription of its own gene.

The EMT model contains eight stable motifs, ranging in size from two to 14 virtual nodes, corresponding to the mesenchymal state and a single, large stable motif corresponding to the epithelial state \citep{steinway_network_2014}. The eight mesenchymal stable motifs overlap each other, for example, the transcription factor SNAI1 participates in five stable motifs. Two of these stable motifs involve the SNAI1-induced production of the signal TGF$\beta$, indicating the self-sustaining nature of TGF$\beta$ signaling.  Panel B of Figure \ref{fig:TLGL&EMT_motifs} illustrates two of the eight stable motifs and their relationships to the marker node E-cadherin. We note that Steinway et al. found that the transcription factor SNAI1 indirectly activates all the other 6 transcription factors (via interactions not shown in the figure), thus its activation is sufficient to downregulate E-cadherin.

\subsection{Domain of influence} 

A key step of the analysis of a Boolean model is determining the influence of a perturbation or permanent intervention on a variable or group of variables. This analysis can be aided by the parity-expanded network, as we describe next following \cite{rozum_parity_2021}, which adapts concepts first introduced by \cite{yang_target_2018}.

Informally, the domain of influence (DOI) describes the set of variable values that will eventually become fixed when an intervention is applied to the system. Formally, we say that a consistent set of virtual nodes $S$, called a seed set, drives a virtual node $I$ if i) $I$ is consistent with $S$ and ii) $F_I (X)=1$ for every attractor state $\boldsymbol{X}$ of the dynamics obtained by restriction to the states in which $S$ is active. Note that this places a condition $F_I(X)=1$ on the update function rather than an apparently simpler condition $\sigma_I(X)=1$ on the variable being updated. The reason for this is that it allows one to consider whether $S$ drives virtual nodes that are themselves members of $S$. The DOI is the set of all virtual nodes driven by $S$ and is written $DOI(S)$. It follows from this definition that (with probability $1$) trajectories with $S$ initially active eventually either inactivate $S$ or activate all of $DOI(S)$. It is often useful to study the subset $DOI(S)-S$ of $DOI(S)$, which consists of all virtual nodes $(i,s)$ not in $S$ for which $X_i=s$ is fixed in all attractors of the dynamics restricted to the subspace defined by $S$. Similarly, $DOI(S)$ in its entirety is the union of $DOI(S)-S$ and the members of $S$ driven by $DOI(S)-S$. We say that $S$ is self-negating if there exists a subset $T$ of $DOI(S)-S$ that drives a virtual node that contradicts $S$. In such cases, $S$ cannot be active in every state of any attractor.

Calculating $DOI(S)$ can be difficult in general, so we focus instead on a commonly used and easily calculated subset of the driving relation: the logical domain of influence (LDOI). A seed set $S$ of virtual nodes logically drives a virtual node $I$ (which may or may not be in $S$) if percolating the fixed variable values defined by $S$ through the update functions results in activating $I$. The set of all $I$ logically driven by $S$ is called the logical domain of influence of $S$, written $LDOI(S)$ (or $LDOI(i,s)$ when $S=\{(i,s)\}$ is of size one). The percolation of variable values can be equivalently expressed as a hyper-path in the parity-expanded network~\citep{yang_target_2018,rozum_parity_2021}. As illustrated on panels C, D of Figure \ref{fig:PEN}, the hyper-path starts with the virtual nodes in $S$ and includes all virtual nodes that are targets of hyperedges that start at element(s) of $S$. For example, in panel C, $S=\{(A,1),(B,1)\}$ and $LDOI(S)=\{(A,1),(B,1),(C,1),(D,1)\}$. We use the notation $Red(G,S)$ for the reduced parity-expanded network that results after restricting the dynamics to the subspace defined by the activity of $S$ and $LDOI(S)$.

As demonstrated by \cite{yang_target_2018}, if $S$ is a subset of $LDOI(S)$, then $LDOI(S)$ contains the virtual nodes of a stable motif. If $LDOI(S)$ contains a stable motif $M$, we say that $S$ (logically) drives $M$. For example, in panel A of Figure \ref{fig:TLGL&EMT_motifs} the LDOI of any of the four virtual nodes of the light red stable motif contains the whole stable motif. Similarly, in the same figure, the LDOI of the ON state of ERK contains the ON state of FasL, but not the ON state of IL2RBT, even though ERK directly activates IL2RBT; this is because the activation of IL2RBT by ERK is reliant upon TBET. On the other hand, the ON state of IL2RBT (and indeed the entire dark blue stable motif and the three uncolored virtual nodes) is contained in the LDOI of the set containing the ON states of both ERK and TBET.

\subsection{Conditionally stable motifs} 
The concept of a conditionally stable motif captures the idea that a normally unstable state of a circuit can become self-sustaining under certain system configurations. More precisely, a \emph{conditionally stable motif} (CSM), $M_c$ of a parity-expanded network, $G$, is a stable motif of a reduced system, $Red(G,C)$, where $C$ is a set of \emph{conditions} for $M_c$ and $Red(G,C)$ is the system obtained by percolating $C$ (plugging in $C$ and its LDOI). To characterize the effects of substituting the conditions $C$ into the system, we consider the hypergraph $G\big |_C$ that describes the update functions of the Boolean network when $C$ is active.
Specifically, let $G\big |_C$ be the hypergraph whose vertex set is $V(G\big |_C)=V(G)$ and whose hyperedge set is formed by removing all parents in $C$ from all hyperedges of $G$; hyperedges without parents remaining are deleted. If the subgraph of $G\big |_C$ induced by the virtual nodes of $M_c$ is strongly connected, then we say that $C$ is a set of \emph{direct} conditions for $M_c$. Otherwise, we say that $C$ is a set of \emph{indirect} conditions for $M_c$. Intuitively, direct conditions are regulators of CSM nodes whose direct effects allow the CSM to become self-sustaining, while indirect conditions enable this self-sustaining behavior through intermediary nodes. If $C$ is a minimal set of direct conditions for $M_c$, i.e., there is no subset of $C$ that is also a set of direct conditions for $M_c$, then we say that $C$ is a set of \emph{strict} requirements for $M_c$.

The T-LGL leukemia model has two conditionally stable motifs. One expresses the activation of the MAPK pathway and the IL2RB receptor protein (virtual nodes with dark blue background in Figure \ref{fig:TLGL&EMT_motifs})A, and its strict requirement is the activation of TBET. The other conditionally stable motif expresses the activity of Ceramide and Fas and the inactivity of S1P and sFas (green background); this motif's strict requirement is the activation of the proteins FasL and NF$\kappa$B, which in turn are in the LDOI of the first (dark blue) conditionally stable motif.

While stable motifs can be thought of as irreversible switches, conditionally stable motifs are irreversible \emph{only if}  their conditions remain satisfied. The same positive circuit of a molecular interaction graph may lead to a stable motif in one cellular context and to a conditionally stable motif in another. For example, a follow-up study of a Boolean model of the mammalian cell cycle~\citep{deritei_principles_2016} found that the cyclin proteins of the phase switch form three stable motifs, which ultimately determine three cell cycle checkpoints. A modified version of the model, in which the checkpoint-determining proteins are disabled, contains only conditionally stable motifs, all of which have the property that their LDOI contains the negation of their conditions. These conditionally stable motifs determine a cyclic behavior that recapitulates the phases of the cell cycle and that is remarkably robust despite the stochastic timing of the model \citep{deritei_feedback_2019}.

\section{State-space compression and attractor identification using stable motifs}
In this section we will describe, in general terms, how to use stable motifs to identify attractors. For additional details about implementation, we refer the reader to \cite{rozum_parity_2021}.

When the variable values corresponding to a stable motif are obtained in the course of a trajectory's time evolution, these variables become fixed for all future times. In particular, these variables are fixed in the attractor to which the trajectory eventually converges.
Thus, the influence of the fixed variables eventually propagates throughout the network, fixing additional variables. Any free variables that remain are governed by update functions that potentially have some of their inputs fixed by the influence of the stable motif.
These update functions define a new, reduced Boolean network that has its own stable motifs, some of which may have been present in the original (unreduced) network, and some of which are created by the variables that are fixed due to the influence of the original stable motif.
This property of the system's natural dynamics enables an iterative reduction approach, in which the effects of activating each stable motif are considered in turn, and at each stage the reduced network induced by each stable motif is constructed. Eventually, the procedure terminates when all possible combinations of stable motif activation have been considered.
Reduced networks without stable motifs are recorded; a subset of the free variables in these reduced networks are guaranteed to oscillate in an attractor of the original system. All possible permutations of stable motif activation are recorded in the succession diagram, which is essentially a permutation tree with nodes corresponding to equivalent reduced networks contracted, and whose leaf nodes present attractors of the original system.

\begin{figure}
    \centering
    \includegraphics[width=1\textwidth]{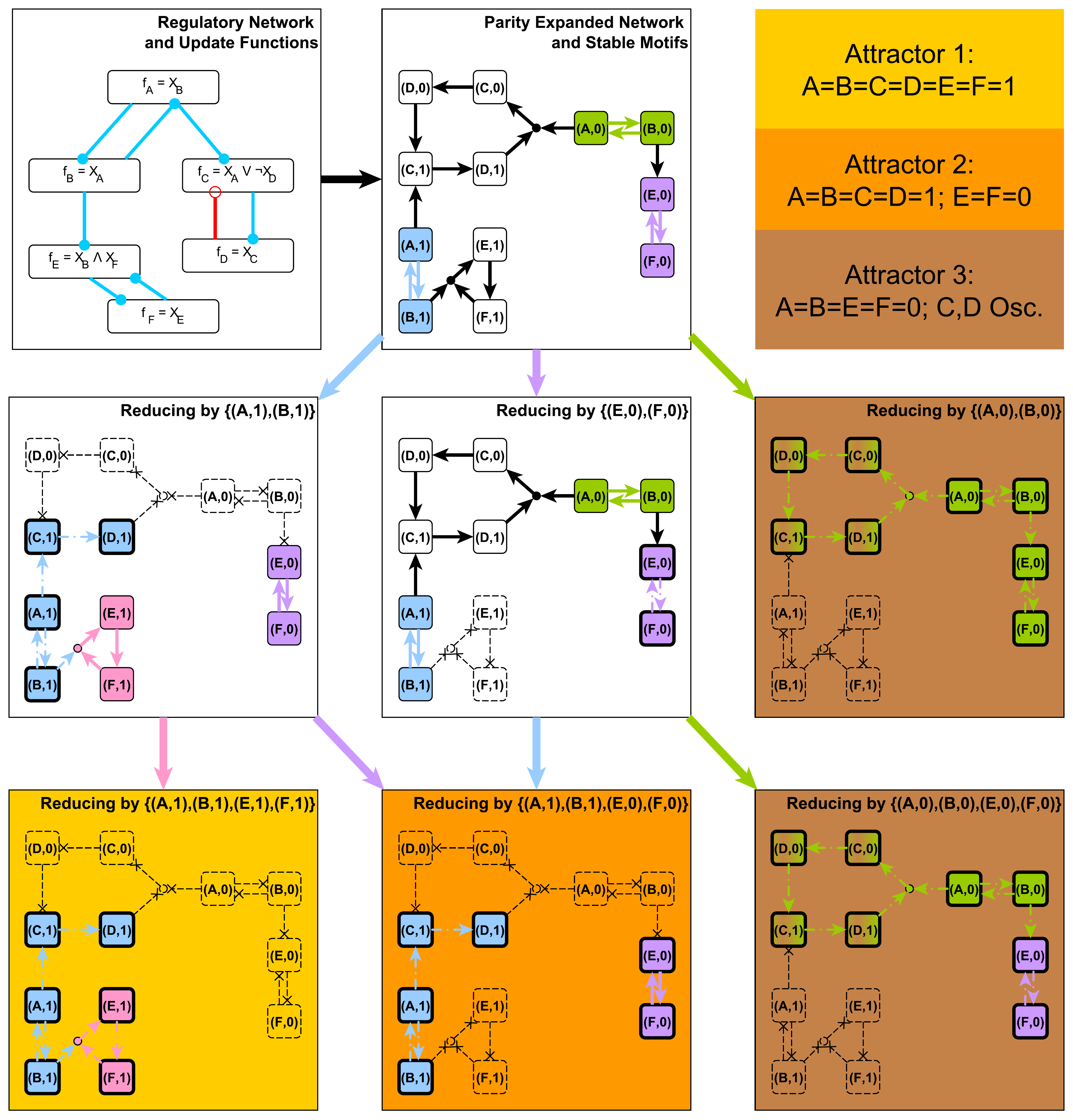}
    \caption{Example of succession diagram construction. The succession diagram is constructed from the Boolean network (top left) translated into $G$, the parity-expanded network (top middle). There are three stable motifs, highlighted in blue, green, and purple. When the the stable motif corresponding to $X_A = X_B = 1$ is selected (following the blue arrow to the middle left panel) its sustained activity leads to $X_C = X_D = 1$ (highlighted in blue). The virtual nodes with dashed outlines become inactive, as their contradictory virtual nodes are locked into the active state. The corresponding reduced system $Red(G, \{(A,1), (B,1)\})$ has two stable motifs, highlighted in pink and purple; the latter was also a stable motif of the original system. Selecting the pink stable motif yields Attractor 1 (highlighted in yellow) and selecting the purple stable motif yields Attractor 2 (orange). In a similar way, each of the other two stable motifs is selected (purple and green arrows, respectively). When the stable motif corresponding to $X_A = X_B = 0$ is selected (green arrow), its sustained activity leads to $X_E=X_F=0$, and the sustained oscillation of nodes C and D, forming Attractor 3 (brown).  }
    \label{fig:sd_attractor_ex}
\end{figure}

The structure of the succession diagram is independent of the updating scheme used \citep{rozum_parity_2021}; however, the relationship between the succession diagram and the repertoire of attractors does depend on the update scheme. In general, at each stage in the iterative process, one must consider whether an attractor may exist that avoids activating any of the remaining stable motifs, thereby prematurely terminating the procedure. Such attractors are called \emph{motif-avoidant}. \cite{rozum_parity_2021} achieve this in the asynchronous case by identifying a series of constraints such an attractor must satisfy, and then using bespoke simulation methods that search the STG for attractors that satisfy these constraints. For example, if a node's activation or oscillation is sufficient to drive (activate) a stable motif, that node, and by extension its update function, must be fixed in its OFF state in any motif-avoidant attractor. In practice, the constraints one can identify will often dramatically reduce the size of the state space one must explore to identify these attractors; indeed, the constraints are often found to be mutually exclusive. Furthermore, it is shown that motif-avoidant attractors are not possible under the most permissive Boolean network (MPBN) updating scheme \citep{pauleve_reconciling_2020}. \cite{naldi_linear_2022} show that a class of Boolean networks, called L-cuttable, do not permit motif-avoidant attractors under asynchronous update.

\begin{figure}
    \centering
    \includegraphics[width=1\textwidth]{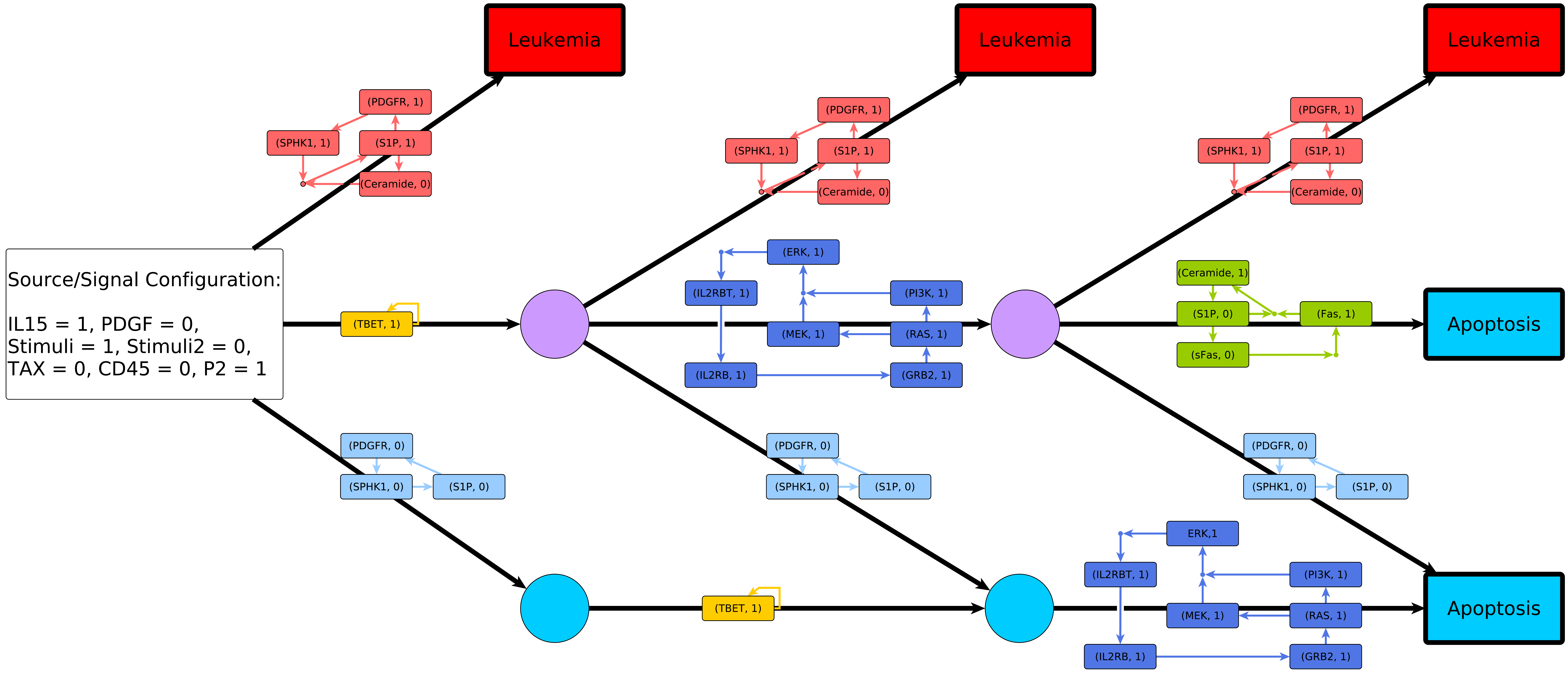}
    \caption{The stable motif succession diagram of the T-LGL network for the specific context (source node configuration) given at the left side of the figure. The root node of the diagram represents the original, unreduced system in the assumed context indicated on that node. The circular symbols represent reduced Boolean networks and the nodes ``Leukemia'' and ``Apoptosis'' refer to attractors representing the eponymous phenotypes. The circular reduced network nodes are colored according to whether both attractors are reachable (purple), only the Leukemia attractor is reachable (red), or only the Apoptosis attractor is reachable (blue) from the trap space corresponding to the reduced network. The edges between the reduced networks correspond to the activation of the stable motif or conditionally stable motif depicted along the edge; these (conditionally) stable motifs are the same as in Figure \ref{fig:TLGL&EMT_motifs}.   }
    \label{fig:TLGL_sd}
\end{figure}

As an example, we look at the succession diagram of the T-LGL network in the context in which antigen stimulation (Stimuli) and the cytokine IL15 are present and the growth factor PDGF is absent. The stable motifs of this network are indicated in \ref{fig:TLGL&EMT_motifs}A.  The stable motif succession diagram contains three main branches, each of which represents the activation of one of the stable motifs of the system (see Figure \ref{fig:TLGL_sd}). The lock-in of the red stable motif leads directly to an attractor of the system, namely the attractor that represents the pathological survival of T cells; we refer to this attractor as ``Leukemia''. The activation of yellow stable motif preserves both the light blue and red stable motifs and satisfies the condition of the dark blue stable motif; this branch of the succession diagram contains three sub-branches, one of which (via the lock-in of the red stable motif) leads to the Leukemia attractor. The rest of the branches of the succession diagram (including the branch that starts with the activation of the light blue stable motif) converge to the Apoptosis attractor. In summary, the natural dynamics of the system reflect a decision between two mutually exclusive stable motifs, which determine two attractors. Knowledge of these decisions can be used to drive the system to a desired attractor, as we will show in the next section.

\subsection{Functional relationships between (conditionally) stable motifs}

\begin{figure}
    \centering
    \includegraphics[width=1\textwidth]{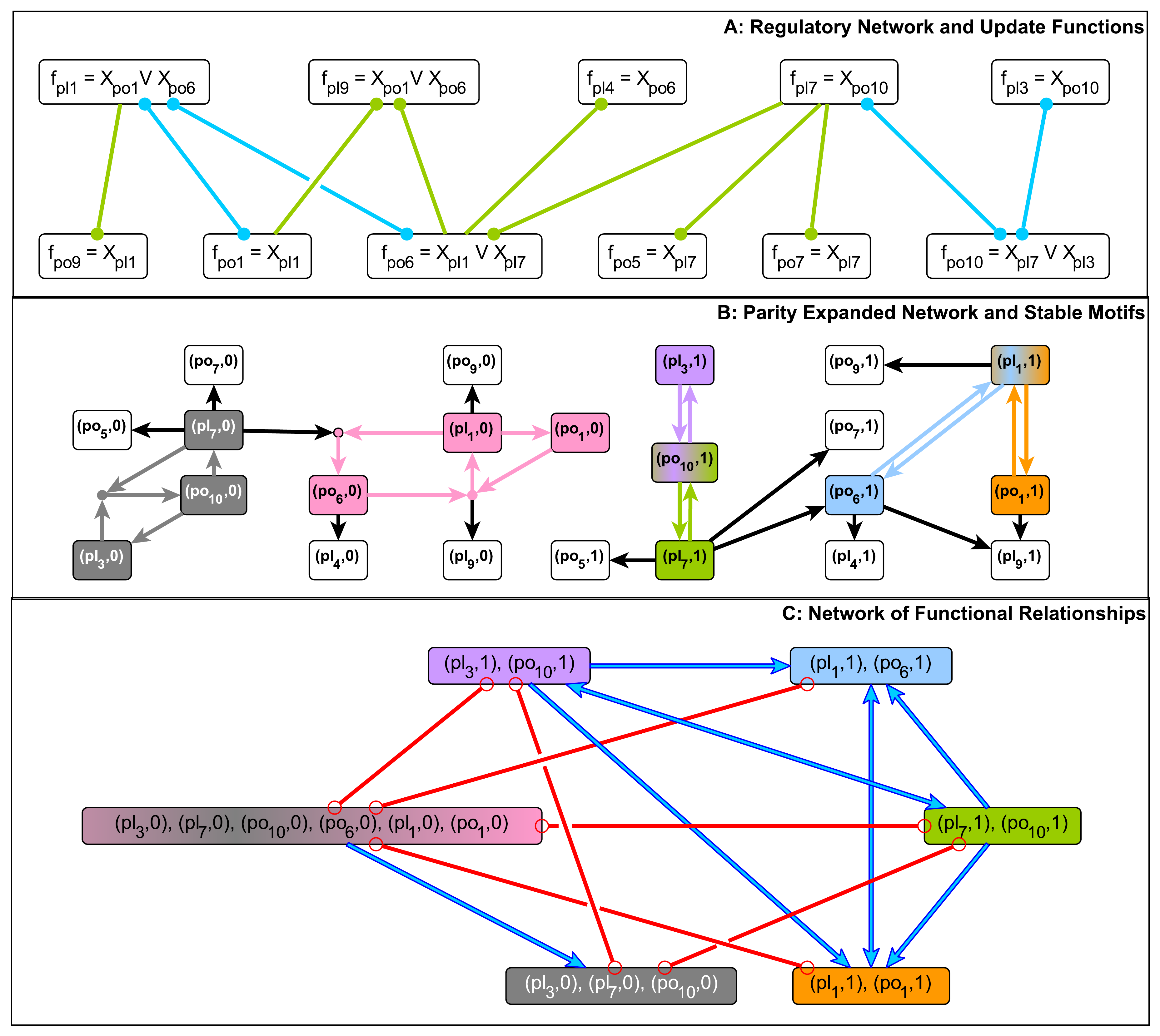}
    \caption{Functional relationships between (conditionally) stable motifs in the Boolean network that corresponds to Figure \ref{fig:ppn_ex}. Panel A indicates the interaction network and the Boolean functions. All the interactions are beneficial; mutually beneficial interactions (bidirectional edges) are shown in blue, while unidirectional edges are shown in green. Panel B illustrates the parity-expanded network for this Boolean system. It contains five stable motifs colored in blue, orange, green, purple, and grey, and one conditionally stable motif highlighted in pink. Panel C shows the network of functional relationships between the stable motifs and the motif group formed by the conditionally stable motif plus its support, the grey stable motif. Blue arrow-tipped edges indicate logical determination and red edges ending in hollow circles represent mutual exclusivity. }
    \label{fig:CSM_relations}
\end{figure}

Each branch of the succession diagram $\Sigma$ contains stable motifs and conditionally stable motifs whose conditions are satisfied by stable motifs that locked in earlier in the succession. These (conditionally) stable motifs can have three types of relationships with each other~\citep{nasrollahi2021relationships}: dependence, mutual exclusivity and logical determination. 
\begin{enumerate}
    \item Dependence: If motif $M_c$ is a conditionally stable motif (CSM) and has a set of conditions $C$ that can be satisfied by motif $M$, then motif $M_c$ is dependent on motif $M$. In this case, motif $M$ is called the support of the conditionally stable motif $M_c$. Dependence means that the support of each CSM must lock in before the CSM does. As a result, it is not possible to have a branch in the succession diagram $\Sigma$ that includes a CSM without its support.
    \item Logical determination: Motif $M_1$ logically determines motif $M_2$ if $M_2 \subset LDOI(M_1)$. If two motifs logically determine each other, they can activate in either order and they will confine the system to the same subspace. 
    \item Mutual exclusivity: Two motifs $M_1$ and $M_2$ are mutually exclusive if they share at least one pair of contradictory virtual nodes (which correspond to different states of the same node of the interaction network). Mutually exclusive motifs cannot appear in the same branch of the succession diagram.
\end{enumerate}

These relationships are stored in the form of a network of functional relationships. We use the example plant-pollinator network of Figure~\ref{fig:ppn_ex} to illustrate this concept. In Figure~\ref{fig:CSM_relations}C, each node is either a stable motif or a motif group. Each motif group consists of one CSM plus one of its supports. This entity encapsulates the first functional relationship, dependence (see the pink-grey motif group). The two remaining functional relationships are represented using edges: Logical determination is shown using blue arrow-tipped edges, e.g., the edges from the green stable motif to the blue, orange, and purple stable motifs demonstrate that if the green stable motif stabilizes, all three will stabilize as well. Mutual exclusivity is shown using the red edges that end in hollow circles, e.g., the edge between the grey and green stable motif represents the fact that these two motifs involve opposite states of the nodes pl\textsubscript{7} and po\textsubscript{10}. This last relationship, mutual exclusivity between the stable motifs and motif groups, has been shown to be useful to efficiently find the number of attractors in plant-pollinator networks. One can combine motifs and motif groups such that each combination is self-consistent (i.e., it does not contain two stable motifs or motif groups that are mutually exclusive) and mutually exclusive with the other combinations. Counting the number of such combinations successfully leads to the number of attractors. In the example of Figure~\ref{fig:CSM_relations} the motif group highlighted with pink and grey in panel C is mutually exclusive with all other stable motifs and its LDOI contains all the virtual nodes in the left connected component of the expanded network shown in panel B.  Hence, this motif group leads to a point attractor of the system. The combination of grey, blue, and orange stable motifs is also consistent and mutually exclusive with the first identified attractor, and consequently leads to the second attractor, also a fixed point. The last consistent combination consists of purple, green, blue, and orange stable motifs, which is mutually exclusive with the first two; this last combination also leads to a distinct point attractor. It was confirmed by other attractor identification methods that this systematic accounting of the functional relationships between stable motifs and motif groups can correctly and efficiently identify the number of attractors in plant-pollinator networks\citep{nasrollahi2021relationships}.

A key motif relationship in the T-LGL leukemia model is the mutual exclusivity between the two stable motifs that contain opposite states of PDGFR, SPHK1 and S1P (light blue and light red nodes in panel A of Figure \ref{fig:TLGL&EMT_motifs}). In any trajectory in which one of these motifs locks in, the other motif cannot ever lock in. Another important motif relationship in this model is the chain of dependence formed between the yellow, dark blue, and green motifs in Figure \ref{fig:TLGL&EMT_motifs}A. The dark blue CSM depends on the active state of TBET (yellow SM), and the green CSM depends on the dark blue CSM. In the EMT network, the sole epithelial stable motif and any individual mesenchymal stable motif (of which there are eight) are mutually exclusive.

\section{Attractor control}
Control theory focuses on developing tools to drive a system to a desired state; this theory usually uses a dynamical model of the system. Network control strategies focus on interacting systems, represented as networks, and encompass a variety of control methods that leverage the properties of the network to reach a target state from a state that does not converge into the target state in the natural dynamics of the system \citep{liu_control_2016,rozum_leveraging_2022}. The most appropriate control methods depend on what the target state is. In some cases, the target state may be any possible state of the system, called full control, whereas in other cases, the target state is one of the dynamical attractors of the model. Here, we focus on this latter case of attractor control.

Attractor control is especially appropriate in biological contexts because biological applications such as cell re-engineering consist of driving to a natural state of the system that would be unreachable without control, e.g., driving from a differentiated cell state to an undifferentiated cell state \citep{zanudo_structure-based_2017}. The goal of these attractor control methods is to identify a set of nodes or edges and the appropriate control action (e.g., permanently deactivating a node) to force the system into the target attractor. Here we will describe two methods with different requirements that can drive the system to a target attractor: feedback vertex set control and succession diagram control. We will also present how these methods can be used to prevent the system from converging into an undesirable attractor.

\subsection{Feedback vertex set control}
\begin{figure}
    \centering
    \includegraphics[width=1\textwidth]{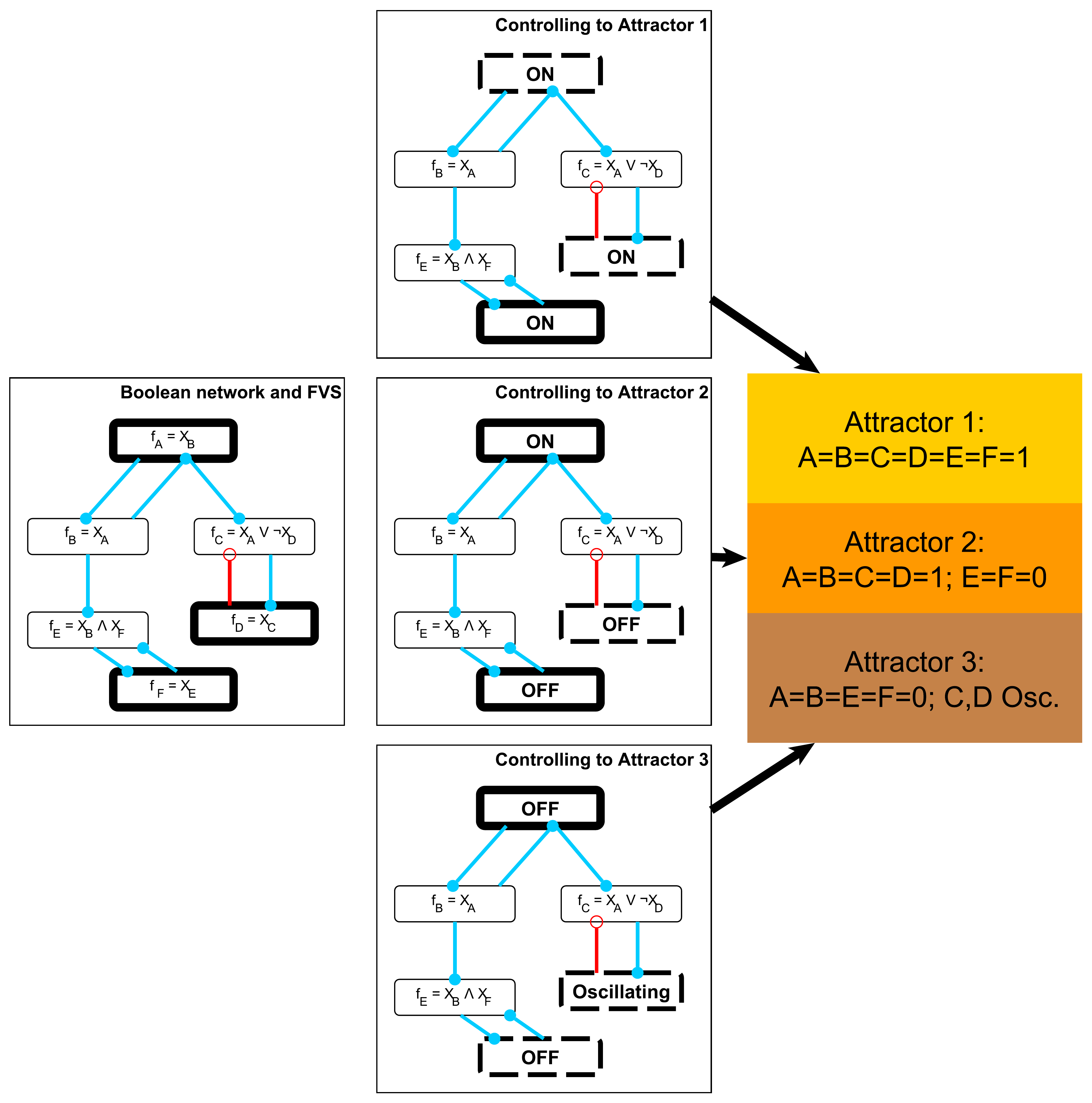}
    \caption{Example of a feedback vertex set and its ability to control attractor dynamics in a Boolean network. The network from Figure \ref{fig:sd_attractor_ex} is shown on the left with a minimal FVS highlighted by bolded outlines, and the attractors of this network are listed on the right. Note that the choice of minimal FVS is not unique. Controlling all nodes in any FVS is sufficient to drive the system to any of its attractors, as is depicted using the chosen FVS in the middle column. Nodes outlined in a dashed line do not need to be overridden to drive the system to the corresponding attractor. In this example, it is never necessary to override node D, though other systems with the same (unsigned) regulatory topology could require control of this node.}
    \label{fig:fvs_ex}
\end{figure}

Feedback vertex set (FVS) control is a powerful technique for driving a dynamical system into one of its attractors \citep{fiedler_dynamics_2013,mochizuki_dynamics_2013, zanudo_structure-based_2017}. A FVS is a set of network nodes whose removal renders the network acyclic. Thus, by controlling the values of the nodes in an FVS (as well as the values of any source nodes), the dynamics of the remaining free variables take place on a network without cycles. Without cycles, a network's dynamics can only have one attractor per input signal combination \citep{snoussi_logical_1993,richard_cycles_2019}. Therefore, controlling the nodes in an FVS along with the input signals guarantees convergence to a unique attractor. If the nodes are controlled to take on values consistent with a target attractor of the uncontrolled system, then the controlled system will converge to this target attractor. Notably, this control strategy does not depend on the initial state of the system. Though this framework was initially developed to study ODEs, the principles apply to Boolean networks as well \citep{zanudo_structure-based_2017}. This is illustrated using the Boolean network of Figure \ref{fig:fvs_ex}. Controlling the values of the three nodes of a minimal FVS (nodes highlighted with bold outlines) can drive this system, and any Boolean network that has the same interaction network, to any of its attractors.

\subsubsection{Feedback vertex subset control}
In biological networks, the minimal FVS is typically larger than the size of the interventions that can be experimentally implemented. Luckily, driving the entire FVS is a sufficient but not necessary condition for driving the network into one of its attractors (as illustrated in Figure \ref{fig:sd_attractor_ex}). The FVS provides an understanding of the system's cycle structure, so it provides a good starting point for identifying attractor driver sets using solely the topology of the system. Various topological metrics can be used in conjunction with the FVS to identify subsets of the FVS that can reliably drive the network.

A class of topological metrics that measure the propagation of information through a network were shown to accurately rank FVS subsets of an ensemble of published Boolean models of biological systems \citep{newby-chaos-2022}. Three propagation metrics: PRINCE propagation \citep{santolini_PNAS_2018}, a modified version of PRINCE propagation \citep{newby-chaos-2022}, and the CheiRank \citep{zhirov_2010} are able to emulate the dynamics of the system. The two PRINCE propagation metrics function by applying a constant perturbation to a set of nodes and measuring how the perturbation propagates through the system. They differ in their normalization of the perturbation: the original PRINCE propagation's normalization is representative of mass-flow, while the modified PRINCE propagation better reflects information-flow. The CheiRank is the PageRank on the network variant in which every edge is reversed, thus it measures the probability of a random walk originating at the focal node.

The combination of these propagation metrics is able to capture the path information and cycle information of the network topology, which, when combined with the cycle information inherent to restricting the subset choice to a FVS subset, is able to accurately estimate the spreading of Boolean dynamics through a system. Single FVS nodes that score well in all three of these propagation metrics are reasonably successful at driving a network towards a desired attractor and very successful at driving a network away from an undesirable attractor. Furthermore, the top ranking FVS subsets of size three dependably drive the network towards one of its attractors. The size of the minimal FVS in these networks ranged between 4 and 17, with a median of 8. In summary, using these topological measures and the FVS, it is possible to find small driver sets for the system's attractors that are more experimentally realizable than full FVS control.

\subsection{Succession diagram control}
\begin{figure}
    \centering
    \includegraphics[width=1\textwidth]{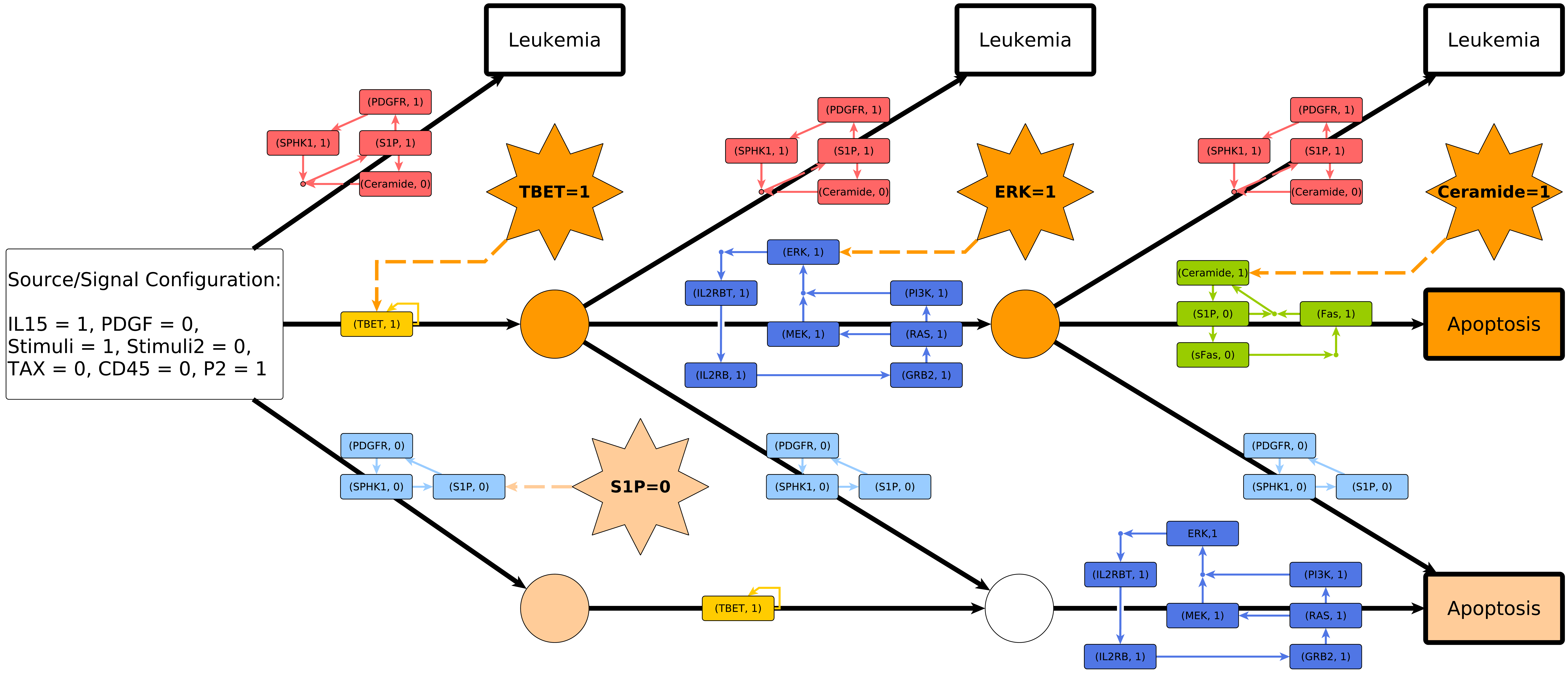}
    \caption{Example of using the succession diagram to drive the T-LGL network to a target attractor. The succession diagram for the T-LGL network model of Figure \ref{fig:TLGL} is shown. The edges of the succession diagram indicate (conditionally) stable motifs that need to be locked in to attain the desired reduced networks, which are represented by the nodes of the succession diagram (circles). These motifs are depicted atop the edges of the succession diagram and are colored as in Figure \ref{fig:TLGL&EMT_motifs}A. Star-shaped nodes indicate control interventions that drive the (conditionally) stable motifs. A target attractor can be reached by selecting a path in the succession diagram from the root node to the target attractor and driving stable motifs along this path. When the system is subjected to the specified control, it will converge to the target attractor from any initial state. Two such control strategies are highlighted here in orange and tan. In the orange control strategy, the middle path in the succession diagram is chosen, and by setting TBET=ERK=Ceramide=1, the three stable motifs that are necessary to attain the target attractor are locked in. In the tan control strategy, only S1P=0 needs to be set. Once the reduced network corresponding to the tan circle is reached, all other paths lead to the apoptosis attractor highlighted in tan.}
    \label{fig:sd_control_ex}
\end{figure}

In Boolean dynamic models, the stable motif succession diagram can be used to identify driver sets for the system's attractors. The general procedure of succession diagram control is to find a minimal driver set for each stable motif that is on the path of the succession diagram leading to the target attractor. On every sequence on the succession diagram leading to the target attractor, locking in each stable motif functions to reduce the state space of the system until the target attractor is the only remaining attractor. Thus, the succession diagram and stable motifs of the system can be utilized to reduce the set of candidates needed to drive the system into one of its attractors (Fig. \ref{fig:sd_control_ex}).

Here, we present the algorithm for succession diagram control to a target attractor $\boldsymbol{X}_D$ developed by \cite{zanudo_cell_2015} and generalized by \cite{rozum_pystablemotifs_2022}. First, using the succession diagram, identify every sequence of (conditionally) stable motifs that ends in attractor $\boldsymbol{X}_D$. For each sequence, remove any inconsequential motifs. Inconsequential stable motifs are the result of a reduced system that only has one accessible stable motif, so they will be locked-in as $t \rightarrow \infty$ and do not need to be driven. Thus, each reduced sequence of stable motifs $\{M_1, M_2,\ldots M_L\}$ provides a framework for what must be driven to lock the system into attractor $\boldsymbol{X}_D$.

For each individual stable motif $M_j$, we find a driver set of virtual nodes $\Delta_j$ such that $M_j \subseteq LDOI(\Delta_j)$. In other words, when fixed, $\Delta_j$ will ensure that the entire stable motif is correctly fixed. Each stable motif is guaranteed to have at least one driver subset (the set $\Delta_j = M_j$); usually smaller driver sets exist. An attractor driver set is found by selecting a driver set for each stable motif in a stable motif sequence that leads to that attractor and taking the union of these selected sets. Because there are often multiple stable motif sequences that lead to the attractor and because each stable motif typically has more than one driver set, the number of attractor driver sets can be large.

The succession diagram control algorithm functions by fixing the node values such that the virtual nodes in each driver set are active. Fixing these node values for a sufficiently long time guarantees that all state transitions eventually lead to states in which every (conditionally) stable motif will lock-in. This intervention can be permanent, but since, by definition, any locked-in stable motif will stay locked-in, the intervention can temporary instead. Thus, succession diagram control is also possible through temporarily fixing each stable motif's driver set until the stable motif is correctly locked-in. This approach can be of use when the driver sets are too large to feasibly fix them simultaneously, or when a driver set of one stable motif is inconsistent with another (this is only possible if drivers are selected from virtual nodes that do not belong to any stable motif in the stable motif sequence).

 This algorithm and multiple variations on it were developed and implemented in the Python library pystablemotifs \citep{rozum_pystablemotifs_2022}; these variations identify different control strategies for the same sequence of stable motifs. The original algorithm presented by \cite{zanudo_cell_2015} searches for internal drivers. That is, for each stable motif only virtual nodes within the stable motif are candidates for driver nodes. This guarantees that a driver set for each stable motif will be efficiently found, but restricting the choice of drivers may prevent the identification of the optimal driver set, as this driver set may involve external drivers (virtual nodes outside of the stable motif). External drivers can reduce the set of drivers because they can drive multiple, logically-independent stable motifs. As an example, the virtual node (B, 0) in Figure~\ref{fig:sd_attractor_ex} is an external driver of the purple stable motif \{(E, 0), (F, 0\}. The network-wide search for external drivers is slower than the local search for internal drivers, but it can be made faster by heuristic node prioritization schemes such as the greedy randomized adaptive search procedure (GRASP) algorithm, which has also been implemented in pystablemotifs\citep{feo_1995,yang_target_2018,rozum_pystablemotifs_2022}. 
 
 Using external drivers can lead to some complications. Because external drivers do not need to be consistent among the stable motifs, it is possible that two contradictory virtual nodes can be among the external drivers for two stable motifs that must both be locked-in, or an external driver of one stable motif may contradict a virtual node in a different target stable motif. To implement such a control strategy, the external driver virtual nodes must be fixed in the correct order based on the succession diagram and they must only be fixed temporarily. These temporarily fixed virtual nodes must be fixed long enough for the stable motif to lock-in before proceeding to the next virtual node, and this process will continue until every stable motif is locked-in, resulting in the desired attractor.

Another variation on this control algorithm, which may reduce the size of driver sets, is the choice of how to pick the targets of the drivers. In the original algorithm, driver sets are found for each stable motif separately, and the attractor driver set is the union of these individual stable motif driver sets. As an alternative, the set of target nodes to be driven can be reduced to a single set: the union of the nodes in every stable motif that must be locked-in. Then, internal or external drivers are found for this single target set instead of for each stable motif individually. Combining the targets in this way presents the same trade-off as using external drivers: the size of the driver set may be smaller, but more computational time is generally required because the size of the state space to be searched is larger. This variation also removes the stable motif lock-in order, which removes the ordering of the driver sets. Without the ordering of the driver sets, it is more difficult to perform succession diagram control through temporarily locking in the driver sets.

As an example, we apply the first algorithm (i.e., finding internal drivers for each stable motif independently) to the T-LGL leukemia model introduced in Section \ref{TLGL}. The succession diagram for the T-LGL model is depicted in Figure \ref{fig:sd_control_ex}, and it shows the two attractors of that model, leukemia and apoptosis, as the possible final states. The figure illustrates two strategies to drive the network into the apoptosis attractor. Using the middle branch of the succession diagram (marked in orange), driving the T-LGL network into the apoptosis attractor needs the lock-in of three (conditionally) stable motifs. Each of these motifs can be driven via a single driver, thus this method indicates that three-node interventions are able to drive the system to the apoptosis state. For example, the three node driving set $\{\textrm{(TBET, 1), (ERK, 1), (Ceramide, 1)}\}$ (indicated by the star-shaped nodes) will lock-in every stable motif on the orange path to apoptosis. Using the lowest branch of the succession diagram (marked in tan), we can also see that driving just (S1P, 0) will also result in the apoptosis attractor. On this branch there is only one possible stable motif in each reduced network. Therefore, the yellow and dark blue stable motifs will automatically activate as the system's dynamics evolve in time (i.e., the yellow and dark blue stable motifs are inconsequential motifs), so these stable motifs do not need to be manually locked-in using additional external controls.

As another example, we consider driving the EMT network introduced in Section \ref{EMT} into the epithelial state from an arbitrary initial state. As the initial state could also be a mesenchymal state, this control encompasses a mesenchymal to epithelial transition. Because the single epithelial stable motif involves the majority of the nodes of the network, a large driver set is necessary to lock-in the stable motif and drive the system into the epithelial attractor\citep{steinway_npjSystemsBio_2015}. There are multiple such five-node sets, including, for example, $\{\textrm{(SHH, 0), (RAS, 0), (MEK, 0), (TGF}\beta\textrm{R, 0), (}\beta\textrm{-catenin}_{\textrm{membrane}}, 1)\}$. It is also of note that every possible five-node driver set consists of virtual nodes whose contradictory partners participate in stable motifs associated with the mesenchymal state.

\subsection{Preventing the system’s convergence into an attractor}
The information incorporated in the succession diagram can also be used to prevent the system from converging into an undesirable (e.g., pathological) attractor. In this case, instead of locking-in the stable motifs that lead to a desired attractor, we block the stable motifs that lead to an undesired attractor. In the following, we present two algorithms for attractor avoidance \citep{zanudo_cell_2015, campbell-chaos-2019} and also present an application to the EMT model first presented in Section \ref{EMT}.

Note that when preventing a biological system from converging to an attractor, it is important to consider the new attractors that may be created by the prevention algorithms. In a biological system the intervention is successful only if any newly created attractors are not biologically similar to the undesired attractor. Verifying this is typically done by looking at biological markers and relies on domain expertise.
Thus, neither of these algorithms may be successful at preventing similar attractors and both must be independently verified.

The first algorithm, introduced by \cite{zanudo_cell_2015} follows a similar procedure to succession diagram control. Given an undesired attractor $\boldsymbol{X}_U$, using the succession diagram, identify the stable motifs that force the system into $\boldsymbol{X}_U$. Each stable motif can be blocked by driving a set of virtual nodes that are the contradictions of some of the stable motif's virtual nodes (i.e., driving a subset of virtual nodes that are part of $\neg M_j$). Driving these virtual nodes will not necessarily fix the entirety of $\neg M_j$, but it will prevent $M_j$ from locking in. Blocking every identified stable motif is not always necessary to prevent the system from converging to $\boldsymbol{X}_U$. After choosing stable motifs to block, the attractor blocking set is found by taking the union of the individual stable motif blocking sets. Fixing the attractor blocking set will prevent the system from converging to $\boldsymbol{X}_U$ and will dramatically reduce the likelihood of converging to biologically similar attractors.

The second algorithm for attractor avoidance using the succession diagram utilizes edgetic perturbations \citep{campbell-chaos-2019}. Edgetic perturbations are perturbations on a single edge of the network, akin to changing a single interaction in a biological network to always be inactive or always be active. These perturbations are more specific than node perturbations, which affect every outgoing edge of the node. In the parity-expanded network, these perturbations manifest as removing parent nodes or child nodes from hyperedges. These topological changes can alter the stable motifs of the system, which can be used to prevent convergence to specific attractors. The attractor prevention algorithm functions by finding the stable motifs that lead to an undesired attractor $\boldsymbol{X}_U$. For each such stable motif, it identifies the largest strongly connected stable module that includes the stable motif, then finds the minimal number of edgetic perturbations that ensures that no subset of the virtual nodes of this stable module forms a stable motif.

Now we will provide an illustrative example of the first attractor prevention method to prevent TGF$\beta$-induced EMT and eventual cancer cell metastasis using the network presented in \ref{EMT}. The convergence of an initially epithelial cell into a mesenchymal attractor can be prevented by blocking the stable motifs that restrict the system into these attractors. In the network, there are seven stable motifs and a stable module composed of a stable motif and a conditionally stable motif that target the mesenchymal state. Locking in any of these eight stable modules guarantees that the unperturbed system will undergo EMT \citep{maheshwari_2017}. While there exist single interventions that prevent the transcriptional downregulation of the marker protein E-cadherin, previous analysis by \cite{steinway_npjSystemsBio_2015} has found that interventions that leave any mesenchymal stable motif intact lead to intermediate cell states that are as pathological, or more so, than mesenchymal cells. To prevent this, we require a set of virtual nodes that blocks all of these stable modules. We find a minimal set of virtual nodes such that there is at least one virtual node from every stable module and then take the contradiction of this set.

One such blocking set is $\{\textrm{(SNAI1, 0), (GLI, 0), (GSK3}\beta,1) \}$. Fixing these virtual nodes in the EMT network does not remove every mesenchymal attractor because the stable motif  $\{\textrm{(Ecadherin, 0),} (\beta\textrm{-catenin}_{\textrm{membrane}}, 0)\}$ still remains in the system. However, this intervention has introduced a new stable motif, $\{\textrm{(Ecadherin,1), }(\beta\textrm{-catenin}_{\textrm{nucleus}},0), (\beta\textrm{-catenin}_{\textrm{membrane}},1)\}$. As this stable motif is active in the epithelial state, EMT can no longer occur after the blocking set is fixed. This blocking set demonstrates the surplus of control in this method because several two-node interventions, such as $\{$(SMAD, 0),(RAS, 0)$\}$, were also experimentally confirmed to suppress TGF$\beta$ induced EMT \citep{steinway_npjSystemsBio_2015}.

\section{Conclusions}

Models of biological systems are useful to the extent that they replicate and predict the empirical behavior of the system being modeled. Because interactions occur at all levels of biological organization and interactions at one level can influence behavior at higher levels, models that can represent diverse biological structures -- especially with varying degrees of empirical characterization -- are of significant interest.      

Indeed, the improvements in experimental technology and the large amounts of generated data have brought the life sciences into an era in which different types of dynamical models are needed to effectively interpret the data and provide system-wide insights into biological systems. Although Boolean models are based on a series of assumptions and have a limited capacity to describe the quantitative features of dynamic systems, here we have shown that they can capture emergent characteristics of real biological systems, demonstrate considerable dynamic richness, and can predict successful therapeutic strategies in biological systems. Boolean network models do not require detailed knowledge of kinetic parameters (as continuous models do), striking a balance between scale and realism. Their parsimonious nature makes them a preferred choice for systems wherein detailed quantitative experimental data is sparse and/or where the system is sufficiently large that continuous models are computationally intractable. 
When additional detail is required, several generalizations are available to consider. For example, while the update functions of Boolean models are usually deterministic and time-independent, it can be beneficial to allow non-deterministic functions, giving rise to a probabilistic Boolean model~\citep{shmulevich_probabilistic_2010}. Such models have multiple update functions for certain nodes and incorporate a probability to switch from one function to the other to better reflect the stochastic aspects of biological regulation, or uncertainty in the implicit cellular context. Studying the deterministic systems that a probabilistic Boolean network switches between using the methods discussed here is a fruitful way to gain insight into the function of the system as a whole.
The success of Boolean networks also indicates that in certain systems the behavior of the system is largely determined by the organization of the network structure rather than the kinetic details of individual interactions; i.e., the dynamic behavior of a network can be encoded in the topology of the parity-expanded network. As a result, the original network's \textit{dynamic} behavior can be efficiently characterized via \textit{structural} analysis. 

A limiting factor to the wide application of Boolean models (and dynamic models in general) is the time and effort required to build and validate the model. The units of the model are the regulatory functions (or differential equations for continuous models) of each node. It is generally the case for each regulatory function that the experimental information available is not sufficient to completely specify the function. Whether constructing the function manually or using inference algorithms (such as \cite{munoz_griffin_2018, wooten_systems-level_2019, razzaq_computational_2018}), modelers need to make assumptions and choices to arrive at a specific function. There isn't a systematic way to test functions one by one; the only way to test them is to test the model as a whole, e.g., by comparing the attractors of the model to experimentally observed phenotypes. Furthermore, in case of discrepancy between model and experimental results, there isn't an unambiguous way of identifying which function needs improvement. In summary, model improvement involves time-consuming trial and error. Automating this process would significantly speed it up. This automation will need to involve the encoding of the multiple types of relevant experimental information into a unified model input format, the quantification of the agreement of model and experimental results, and a process that generates and scores model variants. 

Once a Boolean model has been sufficiently validated, many tools are available for analyzing the emergent properties of the system. In biological systems, the manner in which collective behaviors arise is an important area of study. For example, how do transcriptional events coordinate to give rise to a cellular phenotype? How do interactions among different species give rise to a stable community? Thus, identifying the properties of a system that give rise to its attractors can lend great biological insights. Attractor identification through the state transition graph is a baseline state space analysis for Boolean models, but it is limited by network size and it does not readily provide explanations for the behavior of the system. A more scalable and informative approach is to study the structure of trap spaces, or self-sustaining behaviors of the system. Trap spaces can be identified by finding (conditionally) stable motifs through the parity-expanded network. These are integral to the attractor repertoire of a Boolean system and can be used to efficiently calculate attractors. The power of stable motif analysis is further emphasized by its insight into attractor driver sets. Stable motifs and the succession diagram provide a powerful framework for indicating important subgraphs that determine long-term behaviors, which greatly reduces the search space for drivers. Another tool for driving long-term behaviors is the feedback vertex set, which utilizes a network's cycle structure to reduce multistability. In summary, Boolean networks serve as a useful foundation for modeling biological (e.g., molecular) systems; they can identify the network features that are key determinants of the dynamics (e.g., positive feedback loops that determine stable motifs) and whose detailed modeling would be most fruitful.

\appendix  


\bibliography{refs}

\begin{thebibliography}{96}
\providecommand{\natexlab}[1]{#1}
\providecommand{\url}[1]{\texttt{#1}}
\expandafter\ifx\csname urlstyle\endcsname\relax
  \providecommand{\doi}[1]{doi: #1}\else
  \providecommand{\doi}{doi: \begingroup \urlstyle{rm}\Url}\fi

\bibitem[Hopfield(1982)]{hopfield_neural_1982}
J~J Hopfield.
\newblock Neural networks and physical systems with emergent collective
  computational abilities.
\newblock \emph{Proceedings of the National Academy of Sciences}, 79\penalty0
  (8):\penalty0 2554--2558, April 1982.
\newblock \doi{10.1073/pnas.79.8.2554}.
\newblock URL \url{https://www.pnas.org/doi/abs/10.1073/pnas.79.8.2554}.
\newblock Publisher: Proceedings of the National Academy of Sciences.

\bibitem[Crutchfield(1994)]{crutchfield_calculi_1994}
James~P. Crutchfield.
\newblock The calculi of emergence: computation, dynamics and induction.
\newblock \emph{Physica D: Nonlinear Phenomena}, 75\penalty0 (1):\penalty0
  11--54, August 1994.
\newblock ISSN 0167-2789.
\newblock \doi{10.1016/0167-2789(94)90273-9}.
\newblock URL
  \url{https://www.sciencedirect.com/science/article/pii/0167278994902739}.

\bibitem[Koorehdavoudi and Bogdan(2016)]{koorehdavoudi_statistical_2016}
Hana Koorehdavoudi and Paul Bogdan.
\newblock A {Statistical} {Physics} {Characterization} of the {Complex}
  {Systems} {Dynamics}: {Quantifying} {Complexity} from {Spatio}-{Temporal}
  {Interactions}.
\newblock \emph{Scientific Reports}, 6\penalty0 (1):\penalty0 27602, June 2016.
\newblock ISSN 2045-2322.
\newblock \doi{10.1038/srep27602}.
\newblock URL \url{https://www.nature.com/articles/srep27602}.
\newblock Number: 1 Publisher: Nature Publishing Group.

\bibitem[Kivelson and Kivelson(2016)]{kivelson_defining_2016}
Sophia Kivelson and Steven~A. Kivelson.
\newblock Defining emergence in physics.
\newblock \emph{npj Quantum Materials}, 1\penalty0 (1):\penalty0 1--2, November
  2016.
\newblock ISSN 2397-4648.
\newblock \doi{10.1038/npjquantmats.2016.24}.
\newblock URL \url{https://www.nature.com/articles/npjquantmats201624}.
\newblock Number: 1 Publisher: Nature Publishing Group.

\bibitem[Zheng(2021)]{zheng_introduction_2021}
Zhigang Zheng.
\newblock An {Introduction} to {Emergence} {Dynamics} in {Complex} {Systems}.
\newblock In Xiang-Yang Liu, editor, \emph{Frontiers and {Progress} of
  {Current} {Soft} {Matter} {Research}}, Soft and {Biological} {Matter}, pages
  133--196. Springer, Singapore, 2021.
\newblock ISBN 9789811592973.
\newblock \doi{10.1007/978-981-15-9297-3_4}.
\newblock URL \url{https://doi.org/10.1007/978-981-15-9297-3_4}.

\bibitem[Rämö et~al.(2007)Rämö, Kauffman, Kesseli, and
  Yli-Harja]{ramo_measures_2007}
Pauli Rämö, Stuart Kauffman, Juha Kesseli, and Olli Yli-Harja.
\newblock Measures for information propagation in {Boolean} networks.
\newblock \emph{Physica D: Nonlinear Phenomena}, 227\penalty0 (1):\penalty0
  100--104, March 2007.
\newblock ISSN 0167-2789.
\newblock \doi{10.1016/j.physd.2006.12.005}.
\newblock URL
  \url{https://www.sciencedirect.com/science/article/pii/S0167278906004829}.

\bibitem[Barros(2009)]{barros_information_2009}
João Barros.
\newblock Information {Flows} in {Complex} {Networks}.
\newblock In Frank Emmert-Streib and Matthias Dehmer, editors,
  \emph{Information {Theory} and {Statistical} {Learning}}, pages 267--287.
  Springer US, Boston, MA, 2009.
\newblock ISBN 978-0-387-84816-7.
\newblock \doi{10.1007/978-0-387-84816-7_11}.
\newblock URL \url{https://doi.org/10.1007/978-0-387-84816-7_11}.

\bibitem[Harush and Barzel(2017)]{harush_dynamic_2017}
Uzi Harush and Baruch Barzel.
\newblock Dynamic patterns of information flow in complex networks.
\newblock \emph{Nature Communications}, 8\penalty0 (1):\penalty0 2181, December
  2017.
\newblock ISSN 2041-1723.
\newblock \doi{10.1038/s41467-017-01916-3}.
\newblock URL \url{https://www.nature.com/articles/s41467-017-01916-3}.
\newblock Number: 1 Publisher: Nature Publishing Group.

\bibitem[Maheshwari and Albert(2017)]{maheshwari_2017}
Parul Maheshwari and R{\'e}ka Albert.
\newblock A framework to find the logic backbone of a biological network.
\newblock \emph{BMC Syst Biol}, 11\penalty0 (1):\penalty0 122, December 2017.

\bibitem[Loskot et~al.(2019)Loskot, Atitey, and
  Mihaylova]{loskot_comprehensive_2019}
Pavel Loskot, Komlan Atitey, and Lyudmila Mihaylova.
\newblock Comprehensive review of models and methods for inferences in
  bio-chemical reaction networks.
\newblock \emph{Frontiers in Genetics}, 10, 2019.
\newblock ISSN 1664-8021.
\newblock \doi{10.3389/fgene.2019.00549}.
\newblock URL
  \url{https://www.frontiersin.org/articles/10.3389/fgene.2019.00549}.

\bibitem[von Dassow et~al.(2000)von Dassow, Meir, Munro, and
  Odell]{von_dassow_segment_2000}
George von Dassow, Eli Meir, Edwin~M. Munro, and Garrett~M. Odell.
\newblock The segment polarity network is a robust developmental module.
\newblock \emph{Nature}, 406\penalty0 (6792):\penalty0 188--192, July 2000.
\newblock ISSN 0028-0836, 1476-4687.
\newblock \doi{10.1038/35018085}.
\newblock URL \url{http://www.nature.com/articles/35018085}.

\bibitem[von Dassow and Odell(2002)]{von_dassow_design_2002}
George von Dassow and Garrett~M. Odell.
\newblock Design and constraints of {theDrosophila} segment polarity module:
  {Robust} spatial patterning emerges from intertwined cell state switches.
\newblock \emph{Journal of Experimental Zoology}, 294\penalty0 (3):\penalty0
  179--215, October 2002.
\newblock ISSN 0022-104X, 1097-010X.
\newblock \doi{10.1002/jez.10144}.
\newblock URL \url{http://doi.wiley.com/10.1002/jez.10144}.

\bibitem[Thomas and Kaufman(2001{\natexlab{a}})]{thomas_multistationarity_2001}
Ren\'e' Thomas and Marcelle Kaufman.
\newblock Multistationarity, the basis of cell differentiation and memory.
  {II}. {Logical} analysis of regulatory networks in terms of feedback
  circuits.
\newblock \emph{Chaos: An Interdisciplinary Journal of Nonlinear Science},
  11\penalty0 (1):\penalty0 180--195, March 2001{\natexlab{a}}.
\newblock ISSN 1054-1500.
\newblock \doi{10.1063/1.1349893}.
\newblock URL \url{https://aip.scitation.org/doi/abs/10.1063/1.1349893}.
\newblock Publisher: American Institute of Physics.

\bibitem[Thomas and
  Kaufman(2001{\natexlab{b}})]{thomas_multistationarity_2001-1}
Ren'e Thomas and Marcelle Kaufman.
\newblock Multistationarity, the basis of cell differentiation and memory. {I}.
  {Structural} conditions of multistationarity and other nontrivial behavior.
\newblock \emph{Chaos: An Interdisciplinary Journal of Nonlinear Science},
  11\penalty0 (1):\penalty0 170--179, March 2001{\natexlab{b}}.
\newblock ISSN 1054-1500.
\newblock \doi{10.1063/1.1350439}.
\newblock URL \url{https://aip.scitation.org/doi/abs/10.1063/1.1350439}.
\newblock Publisher: American Institute of Physics.

\bibitem[Rozum and Albert(2018{\natexlab{a}})]{rozum_identifying_2018}
Jordan~C. Rozum and Réka Albert.
\newblock Identifying (un)controllable dynamical behavior in complex networks.
\newblock \emph{PLOS Computational Biology}, 14\penalty0 (12):\penalty0
  e1006630, December 2018{\natexlab{a}}.
\newblock ISSN 1553-7358.
\newblock \doi{10.1371/journal.pcbi.1006630}.
\newblock URL \url{https://dx.plos.org/10.1371/journal.pcbi.1006630}.

\bibitem[Rozum and Albert(2018{\natexlab{b}})]{rozum_self-sustaining_2018}
Jordan~C. Rozum and Réka Albert.
\newblock Self-sustaining positive feedback loops in discrete and continuous
  systems.
\newblock \emph{Journal of Theoretical Biology}, 459:\penalty0 36--44, December
  2018{\natexlab{b}}.
\newblock ISSN 00225193.
\newblock \doi{10.1016/j.jtbi.2018.09.017}.
\newblock URL
  \url{https://linkinghub.elsevier.com/retrieve/pii/S0022519318304533}.

\bibitem[Rozum and Albert(2019)]{rozum_controlling_2019}
Jordan~C. Rozum and Réka Albert.
\newblock Controlling the cell cycle restriction switch across the information
  gradient.
\newblock \emph{Advances in Complex Systems}, 22\penalty0 (07n08):\penalty0
  1950020, November 2019.
\newblock ISSN 0219-5259.
\newblock \doi{10.1142/S0219525919500206}.
\newblock URL
  \url{https://www.worldscientific.com/doi/abs/10.1142/S0219525919500206}.
\newblock Publisher: World Scientific Publishing Co.

\bibitem[Liu and Barabási(2016)]{liu_control_2016}
Yang-Yu Liu and Albert-Laszló Barabási.
\newblock Control {Principles} of {Complex} {Networks}.
\newblock \emph{Reviews of Modern Physics}, 88\penalty0 (3):\penalty0 035006,
  September 2016.
\newblock ISSN 0034-6861, 1539-0756.
\newblock \doi{10.1103/RevModPhys.88.035006}.
\newblock URL \url{http://arxiv.org/abs/1508.05384}.
\newblock arXiv: 1508.05384.

\bibitem[Rozum and Albert(2022)]{rozum_leveraging_2022}
Jordan~C. Rozum and Réka Albert.
\newblock Leveraging network structure in nonlinear control.
\newblock \emph{npj Systems Biology and Applications}, 8\penalty0 (1):\penalty0
  38, 2022.
\newblock \doi{10.1038/s41540-022-00249-2}.

\bibitem[Mackey et~al.(2016)Mackey, Santill{\'a}n, Tyran-Kami{\'n}ska, and
  Zeron]{mackey_simple_2016}
Michael~C. Mackey, Moisés Santill{\'a}n, Marta Tyran-Kami{\'n}ska, and
  Eduardo~S. Zeron.
\newblock \emph{Simple {Mathematical} {Models} of {Gene} {Regulatory}
  {Dynamics}}.
\newblock Lecture {Notes} on {Mathematical} {Modelling} in the {Life}
  {Sciences}. Springer International Publishing, Cham, 2016.
\newblock ISBN 978-3-319-45317-0 978-3-319-45318-7.
\newblock \doi{10.1007/978-3-319-45318-7}.
\newblock URL \url{http://link.springer.com/10.1007/978-3-319-45318-7}.

\bibitem[Albert and Thakar(2014)]{albert_boolean_2014}
R\'{e}ka Albert and Juilee Thakar.
\newblock Boolean modeling: a logic-based dynamic approach for understanding
  signaling and regulatory networks and for making useful predictions.
\newblock \emph{WIREs Systems Biology and Medicine}, 6\penalty0 (5):\penalty0
  353--369, 2014.
\newblock \doi{https://doi.org/10.1002/wsbm.1273}.
\newblock URL
  \url{https://wires.onlinelibrary.wiley.com/doi/abs/10.1002/wsbm.1273}.

\bibitem[Abou-Jaoud\'{e} et~al.(2016)Abou-Jaoud\'{e}, Traynard, Monteiro,
  Saez-Rodriguez, Helikar, Thieffry, and Chaouiya]{abou-jaoude_logical_2016}
Wasim Abou-Jaoud\'{e}, Pauline Traynard, Pedro Monteiro, Julio Saez-Rodriguez,
  Tomas Helikar, Denis Thieffry, and Claudine Chaouiya.
\newblock Logical modeling and dynamical analysis of cellular networks.
\newblock \emph{Front Genet}, 7:\penalty0 94, 2016.
\newblock \doi{https://doi.org/10.3389/fgene.2016.00094}.

\bibitem[Schwab et~al.(2020)Schwab, K\"{u}lwein, Ikonomi, K\"{u}hl, and
  Kestler]{schwab_computational_2020}
Julian~D. Schwab, Silke~D. K\"{u}lwein, Nensi Ikonomi, Michael K\"{u}hl, and
  Hans~A. Kestler.
\newblock Concepts in boolean network modeling: What do they all mean?
\newblock \emph{Computational and Structural Biotechnology Journal},
  18:\penalty0 571--582, 2020.
\newblock ISSN 2001-0370.
\newblock \doi{https://doi.org/10.1016/j.csbj.2020.03.001}.
\newblock URL
  \url{https://www.sciencedirect.com/science/article/pii/S200103701930460X}.

\bibitem[Hemedan et~al.(2022)Hemedan, Niarakis, Schneider, and
  Ostaszewski]{hemedan_computational_2022}
Ahmed~Abdelmonem Hemedan, Anna Niarakis, Reinhard Schneider, and Marek
  Ostaszewski.
\newblock Boolean modelling as a logic-based dynamic approach in systems
  medicine.
\newblock \emph{Computational and Structural Biotechnology Journal},
  20:\penalty0 3161--3172, 2022.
\newblock ISSN 2001-0370.
\newblock \doi{https://doi.org/10.1016/j.csbj.2022.06.035}.
\newblock URL
  \url{https://www.sciencedirect.com/science/article/pii/S2001037022002495}.

\bibitem[Alon(2019)]{alon2019}
Uri Alon.
\newblock \emph{An Introduction to Systems Biology: Design Principles of
  Biological Circuits}.
\newblock Chapman \& Hall, 2019.

\bibitem[Saadatpour et~al.(2010)Saadatpour, Albert, and
  Albert]{saadatpour_attractor_2010}
Assieh Saadatpour, Istv\'an Albert, and R\'eka Albert.
\newblock Attractor analysis of asynchronous boolean models of signal
  transduction networks.
\newblock \emph{Journal of Theoretical Biology}, 266\penalty0 (4):\penalty0
  641--656, 2010.
\newblock ISSN 0022-5193.
\newblock \doi{https://doi.org/10.1016/j.jtbi.2010.07.022}.
\newblock URL
  \url{https://www.sciencedirect.com/science/article/pii/S0022519310003796}.

\bibitem[Thomas(1973)]{thomas_boolean_1973}
Ren{\'e} Thomas.
\newblock Boolean formalization of genetic control circuits.
\newblock \emph{Journal of Theoretical Biology}, 42\penalty0 (3):\penalty0
  563--585, 1973.
\newblock ISSN 0022-5193.
\newblock \doi{https://doi.org/10.1016/0022-5193(73)90247-6}.
\newblock URL
  \url{https://www.sciencedirect.com/science/article/pii/0022519373902476}.

\bibitem[Kauffman(1969)]{kauffman_metabolic_1969}
Stuart~A. Kauffman.
\newblock Metabolic stability and epigenesis in randomly constructed genetic
  nets.
\newblock \emph{Journal of Theoretical Biology}, 22\penalty0 (3):\penalty0
  437--467, March 1969.
\newblock ISSN 0022-5193.
\newblock \doi{10.1016/0022-5193(69)90015-0}.
\newblock URL
  \url{https://www.sciencedirect.com/science/article/pii/0022519369900150}.

\bibitem[Aldana and Cluzel(2003)]{aldana_natural_2003}
Maximino Aldana and Philippe Cluzel.
\newblock A natural class of robust networks.
\newblock \emph{Proceedings of the National Academy of Sciences}, 100\penalty0
  (15):\penalty0 8710--8714, 2003.
\newblock \doi{10.1073/pnas.1536783100}.
\newblock URL \url{https://www.pnas.org/doi/abs/10.1073/pnas.1536783100}.

\bibitem[Harris et~al.(2002)Harris, Sawhill, Wuensche, and
  Kauffman]{harris_model_2002}
Stephen~E. Harris, Bruce~K. Sawhill, Andrew Wuensche, and Stuart Kauffman.
\newblock A model of transcriptional regulatory networks based on biases in the
  observed regulation rules.
\newblock \emph{Complexity}, 7\penalty0 (4):\penalty0 23--40, 2002.
\newblock \doi{https://doi.org/10.1002/cplx.10022}.
\newblock URL \url{https://onlinelibrary.wiley.com/doi/abs/10.1002/cplx.10022}.

\bibitem[Raeymaekers(2002)]{raeymaekers_dynamics_2002}
Luc Raeymaekers.
\newblock Dynamics of boolean networks controlled by biologically meaningful
  functions.
\newblock \emph{Journal of Theoretical Biology}, 218\penalty0 (3):\penalty0
  331--341, 2002.
\newblock \doi{10.1006/jtbi.2002.3081}.

\bibitem[Subbaroyan et~al.(2022)Subbaroyan, Martin, and
  Samal]{subbaroyan_minimum_2002}
Ajay Subbaroyan, Olivier~C Martin, and Areejit Samal.
\newblock {Minimum complexity drives regulatory logic in Boolean models of
  living systems}.
\newblock \emph{PNAS Nexus}, 1\penalty0 (1), 04 2022.
\newblock ISSN 2752-6542.
\newblock \doi{10.1093/pnasnexus/pgac017}.
\newblock URL \url{https://doi.org/10.1093/pnasnexus/pgac017}.
\newblock pgac017.

\bibitem[Jarrah et~al.(2007)Jarrah, Raposa, and
  Laubenbacher]{jarrah_nested_2007}
Abdul~Salam Jarrah, Blessilda Raposa, and Reinhard Laubenbacher.
\newblock Nested canalyzing, unate cascade, and polynomial functions.
\newblock \emph{Physica D: Nonlinear Phenomena}, 233\penalty0 (2):\penalty0
  167--174, 2007.
\newblock ISSN 0167-2789.
\newblock \doi{https://doi.org/10.1016/j.physd.2007.06.022}.
\newblock URL
  \url{https://www.sciencedirect.com/science/article/pii/S0167278907002035}.

\bibitem[Bornholdt and Kauffman(2019)]{bornholdt_ensembles_2019}
Stefan Bornholdt and Stuart Kauffman.
\newblock Ensembles, dynamics, and cell types: Revisiting the statistical
  mechanics perspective on cellular regulation.
\newblock \emph{Journal of Theoretical Biology}, 467:\penalty0 15--22, 2019.
\newblock ISSN 0022-5193.
\newblock \doi{https://doi.org/10.1016/j.jtbi.2019.01.036}.
\newblock URL
  \url{https://www.sciencedirect.com/science/article/pii/S0022519319300530}.

\bibitem[Naldi et~al.(2011)Naldi, Remy, Thieffry, and
  Chaouiya]{naldi_dynamically_2011}
Aurélien Naldi, Elisabeth Remy, Denis Thieffry, and Claudine Chaouiya.
\newblock Dynamically consistent reduction of logical regulatory graphs.
\newblock \emph{Theoretical Computer Science}, 412\penalty0 (21):\penalty0
  2207--2218, May 2011.
\newblock ISSN 03043975.
\newblock \doi{10.1016/j.tcs.2010.10.021}.
\newblock URL
  \url{https://linkinghub.elsevier.com/retrieve/pii/S0304397510005839}.

\bibitem[Naldi et~al.(2022)Naldi, Richard, and Tonello]{naldi_linear_2022}
Aur{\'e}lien Naldi, Adrien Richard, and Elisa Tonello.
\newblock Linear cuts in boolean networks, 2022.
\newblock URL \url{https://arxiv.org/abs/2203.01620}.

\bibitem[Li et~al.(2006)Li, Assmann, and Albert]{li_predicting_2006}
Song Li, Sarah~M Assmann, and Réka Albert.
\newblock Predicting {Essential} {Components} of {Signal} {Transduction}
  {Networks}: {A} {Dynamic} {Model} of {Guard} {Cell} {Abscisic} {Acid}
  {Signaling}.
\newblock \emph{PLoS Biology}, 4\penalty0 (10):\penalty0 e312, September 2006.
\newblock ISSN 1545-7885.
\newblock \doi{10.1371/journal.pbio.0040312}.
\newblock URL \url{https://dx.plos.org/10.1371/journal.pbio.0040312}.

\bibitem[Maheshwari et~al.(2022)Maheshwari, Assmann, and
  Albert]{maheshwari_inference_2022}
Parul Maheshwari, Sarah~M. Assmann, and Reka Albert.
\newblock Inference of a boolean network from causal logic implications.
\newblock \emph{Frontiers in Genetics}, 13, 2022.
\newblock ISSN 1664-8021.
\newblock \doi{10.3389/fgene.2022.836856}.
\newblock URL
  \url{https://www.frontiersin.org/articles/10.3389/fgene.2022.836856}.

\bibitem[Razzaq et~al.(2018)Razzaq, Paulev{\'e}, Siegel, Saez-Rodriguez,
  Bourdon, and Guziolowski]{razzaq_computational_2018}
Misbah Razzaq, Loïc Paulev{\'e}, Anne Siegel, Julio Saez-Rodriguez,
  J{\'e}r{\'e}mie Bourdon, and Carito Guziolowski.
\newblock Computational discovery of dynamic cell line specific boolean
  networks from multiplex time-course data.
\newblock \emph{PLOS Computational Biology}, 14\penalty0 (10):\penalty0 1--23,
  10 2018.
\newblock \doi{10.1371/journal.pcbi.1006538}.
\newblock URL \url{https://doi.org/10.1371/journal.pcbi.1006538}.

\bibitem[Mu{\~{n}}oz et~al.(2018)Mu{\~{n}}oz, Carrillo, Azpeitia, and
  Rosenblueth]{munoz_griffin_2018}
Stalin Mu{\~{n}}oz, Miguel Carrillo, Eugenio Azpeitia, and David~A.
  Rosenblueth.
\newblock Griffin: A tool for symbolic inference of synchronous boolean
  molecular networks.
\newblock \emph{Frontiers in Genetics}, 9, 2018.
\newblock ISSN 1664-8021.
\newblock \doi{10.3389/fgene.2018.00039}.
\newblock URL
  \url{https://www.frontiersin.org/articles/10.3389/fgene.2018.00039}.

\bibitem[Saint-Antoine and Singh(2020)]{saint-antoine_network_2020}
MM~Saint-Antoine and A~Singh.
\newblock Network inference in systems biology: recent developments,
  challenges, and applications.
\newblock \emph{Curr Opin Biotechnol.}, 63:\penalty0 89--98, 2020.
\newblock \doi{10.1016/j.copbio.2019.12.002}.

\bibitem[Magrans~de Abril et~al.(2018)Magrans~de Abril, Yoshimoto, and
  Doya]{magrans_connectivity_2018}
Ildefons Magrans~de Abril, Junichiro Yoshimoto, and Kenji Doya.
\newblock Connectivity inference from neural recording data: Challenges,
  mathematical bases and research directions.
\newblock \emph{Neural Networks}, 102:\penalty0 120--137, 2018.
\newblock ISSN 0893-6080.
\newblock \doi{https://doi.org/10.1016/j.neunet.2018.02.016}.
\newblock URL
  \url{https://www.sciencedirect.com/science/article/pii/S0893608018300704}.

\bibitem[Pandey et~al.(2010)Pandey, Wang, Wilson, Li, Zhao, Gookin, Assmann,
  and Albert]{pandey_Boolean_2010}
Sona Pandey, Rui-Sheng Wang, Liza Wilson, Song Li, Zhixin Zhao, Timothy~E
  Gookin, Sarah~M Assmann, and R{\'e}ka Albert.
\newblock Boolean modeling of transcriptome data reveals novel modes of
  heterotrimeric g-protein action.
\newblock \emph{Molecular Systems Biology}, 6\penalty0 (1):\penalty0 372, 2010.
\newblock \doi{https://doi.org/10.1038/msb.2010.28}.
\newblock URL \url{https://www.embopress.org/doi/abs/10.1038/msb.2010.28}.

\bibitem[Wooten et~al.(2019)Wooten, Groves, Tyson, Liu, Lim, Albert, Lopez,
  Sage, and Quaranta]{wooten_systems-level_2019}
David~J. Wooten, Sarah~M. Groves, Darren~R. Tyson, Qi~Liu, Jing~S. Lim, Réka
  Albert, Carlos~F. Lopez, Julien Sage, and Vito Quaranta.
\newblock Systems-level network modeling of {Small} {Cell} {Lung} {Cancer}
  subtypes identifies master regulators and destabilizers.
\newblock \emph{PLOS Computational Biology}, 15\penalty0 (10):\penalty0
  e1007343, October 2019.
\newblock ISSN 1553-7358.
\newblock \doi{10.1371/journal.pcbi.1007343}.
\newblock URL
  \url{https://journals.plos.org/ploscompbiol/article?id=10.1371/journal.pcbi.1007343}.
\newblock Publisher: Public Library of Science.

\bibitem[Albert et~al.(2017)Albert, Acharya, Jeon, Za{\~{n}}udo, Zhu, Osman,
  and Assmann]{albert_new_2017}
R{\'e}ka Albert, Biswa~R. Acharya, Byeong~Wook Jeon, Jorge G.~T. Za{\~{n}}udo,
  Mengmeng Zhu, Karim Osman, and Sarah~M. Assmann.
\newblock A new discrete dynamic model of aba-induced stomatal closure predicts
  key feedback loops.
\newblock \emph{PLOS Biology}, 15\penalty0 (9):\penalty0 1--35, 09 2017.
\newblock \doi{10.1371/journal.pbio.2003451}.
\newblock URL \url{https://doi.org/10.1371/journal.pbio.2003451}.

\bibitem[Maheshwari et~al.(2019)Maheshwari, Du, Sheen, Assmann, and
  Albert]{maheshwari_model_driven_2019}
Parul Maheshwari, Hao Du, Jen Sheen, Sarah~M. Assmann, and Reka Albert.
\newblock Model-driven discovery of calcium-related protein-phosphatase
  inhibition in plant guard cell signaling.
\newblock \emph{PLOS Computational Biology}, 15\penalty0 (10):\penalty0 1--28,
  10 2019.
\newblock \doi{10.1371/journal.pcbi.1007429}.
\newblock URL \url{https://doi.org/10.1371/journal.pcbi.1007429}.

\bibitem[Naldi et~al.(2018{\natexlab{a}})Naldi, Hernandez, Abou-Jaoudé,
  Monteiro, Chaouiya, and Thieffry]{naldi_logical_2018}
Aurélien Naldi, Céline Hernandez, Wassim Abou-Jaoudé, Pedro~T. Monteiro,
  Claudine Chaouiya, and Denis Thieffry.
\newblock Logical {Modeling} and {Analysis} of {Cellular} {Regulatory}
  {Networks} {With} {GINsim} 3.0.
\newblock \emph{Frontiers in Physiology}, 9, 2018{\natexlab{a}}.
\newblock ISSN 1664-042X.
\newblock \doi{10.3389/fphys.2018.00646}.
\newblock URL
  \url{https://www.frontiersin.org/articles/10.3389/fphys.2018.00646/full}.

\bibitem[Correia et~al.(2018)Correia, Gates, Wang, and
  Rocha]{correia_cana_2018}
Rion~B. Correia, Alexander~J. Gates, Xuan Wang, and Luis~M. Rocha.
\newblock {CANA}: {A} {Python} {Package} for {Quantifying} {Control} and
  {Canalization} in {Boolean} {Networks}.
\newblock \emph{Frontiers in Physiology}, 9, 2018.
\newblock ISSN 1664-042X.
\newblock URL
  \url{https://www.frontiersin.org/articles/10.3389/fphys.2018.01046}.

\bibitem[Rozum et~al.(2022)Rozum, Deritei, Park, Gómez Tejeda~Zañudo, and
  Albert]{rozum_pystablemotifs_2022}
Jordan~C Rozum, Dávid Deritei, Kyu~Hyong Park, Jorge Gómez Tejeda~Zañudo,
  and Réka Albert.
\newblock pystablemotifs: {Python} library for attractor identification and
  control in {Boolean} networks.
\newblock \emph{Bioinformatics}, 38\penalty0 (5):\penalty0 1465--1466, March
  2022.
\newblock ISSN 1367-4803.
\newblock \doi{10.1093/bioinformatics/btab825}.
\newblock URL \url{https://doi.org/10.1093/bioinformatics/btab825}.

\bibitem[Naldi et~al.(2018{\natexlab{b}})Naldi, Hernandez, Levy, Stoll,
  Monteiro, Chaouiya, Helikar, Zinovyev, Calzone, Cohen-Boulakia, Thieffry, and
  Paulevé]{naldi_colomoto_2018}
Aurélien Naldi, Céline Hernandez, Nicolas Levy, Gautier Stoll, Pedro~T.
  Monteiro, Claudine Chaouiya, Tomáš Helikar, Andrei Zinovyev, Laurence
  Calzone, Sarah Cohen-Boulakia, Denis Thieffry, and Loïc Paulevé.
\newblock The {CoLoMoTo} {Interactive} {Notebook}: {Accessible} and
  {Reproducible} {Computational} {Analyses} for {Qualitative} {Biological}
  {Networks}.
\newblock \emph{Frontiers in Physiology}, 9:\penalty0 680, June
  2018{\natexlab{b}}.
\newblock ISSN 1664-042X.
\newblock \doi{10.3389/fphys.2018.00680}.
\newblock URL
  \url{https://www.frontiersin.org/article/10.3389/fphys.2018.00680/full}.

\bibitem[Memmott(1999)]{memmott_structure_1999}
Jane Memmott.
\newblock The structure of a plant-pollinator food web.
\newblock \emph{Ecology Letters}, 2\penalty0 (5):\penalty0 276--280, 1999.
\newblock \doi{https://doi.org/10.1046/j.1461-0248.1999.00087.x}.
\newblock URL
  \url{https://onlinelibrary.wiley.com/doi/abs/10.1046/j.1461-0248.1999.00087.x}.

\bibitem[Russo et~al.(2019)Russo, Albert, Campbell, and
  Shea]{russo_experimental_2019}
Laura Russo, R\'{e}ka Albert, Colin Campbell, and Katriona Shea.
\newblock Experimental species introduction shapes network interactions in a
  plant-pollinator community.
\newblock \emph{Biological Invasions}, 21:\penalty0 3505–3519, 2019.
\newblock \doi{https://doi.org/10.1007/s10530-019-02064-z}.

\bibitem[Campbell et~al.(2011)Campbell, Yang, Albert, and
  Shea]{campbell_network_2011}
Colin Campbell, Suann Yang, Réka Albert, and Katriona Shea.
\newblock A network model for plant–pollinator community assembly.
\newblock \emph{Proceedings of the National Academy of Sciences}, 108\penalty0
  (1):\penalty0 197--202, 2011.
\newblock ISSN 0027-8424, 1091-6490.
\newblock \doi{10.1073/pnas.1008204108}.
\newblock URL \url{http://www.pnas.org/content/108/1/197}.

\bibitem[Bascompte et~al.(2003)Bascompte, Jordano, Meli{\'a}n, and
  Olesen]{bascompte2003nested}
Jordi Bascompte, Pedro Jordano, Carlos~J Meli{\'a}n, and Jens~M Olesen.
\newblock The nested assembly of plant--animal mutualistic networks.
\newblock \emph{Proceedings of the National Academy of Sciences}, 100\penalty0
  (16):\penalty0 9383--9387, 2003.

\bibitem[Campbell et~al.(2022{\natexlab{a}})Campbell, Russo, Albert, Buckling,
  and Shea]{campbell_whole_2022}
Colin Campbell, Laura Russo, Réka Albert, Angus Buckling, and Katriona Shea.
\newblock Whole community invasions and the integration of novel ecosystems.
\newblock \emph{{PLOS} Computational Biology}, 18\penalty0 (6):\penalty0
  e1010151, 2022{\natexlab{a}}.
\newblock ISSN 1553-7358.
\newblock \doi{10.1371/journal.pcbi.1010151}.
\newblock URL \url{https://dx.plos.org/10.1371/journal.pcbi.1010151}.

\bibitem[Russo et~al.(2013)Russo, DeBarros, Yang, Shea, and
  Mortensen]{russo_supporting_2013}
Laura Russo, Nelson DeBarros, Suann Yang, Katriona Shea, and David Mortensen.
\newblock Supporting crop pollinators with floral resources: network-based
  phenological matching.
\newblock \emph{Ecology and Evolution}, 3\penalty0 (9):\penalty0 3125--3140,
  2013.
\newblock \doi{https://doi.org/10.1002/ece3.703}.
\newblock URL \url{https://onlinelibrary.wiley.com/doi/abs/10.1002/ece3.703}.

\bibitem[Campbell et~al.(2012)Campbell, Yang, Shea, and
  Albert]{campbell2012topology}
Colin Campbell, Suann Yang, Katriona Shea, and R{\'e}ka Albert.
\newblock Topology of plant-pollinator networks that are vulnerable to collapse
  from species extinction.
\newblock \emph{Physical Review E}, 86\penalty0 (2):\penalty0 021924, 2012.

\bibitem[Allesina and Tang(2012)]{allesina_stability_2012}
Stefano Allesina and Si~Tang.
\newblock Stability criteria for complex ecosystems.
\newblock \emph{Nature}, 483:\penalty0 205--208, 2012.
\newblock \doi{https://doi.org/10.1038/nature10832}.
\newblock URL \url{https://www.nature.com/articles/nature10832}.

\bibitem[Staniczenko et~al.(2013)Staniczenko, Kopp, and
  Allesina]{staniczenko_ghost_2013}
Phillip~P.A. Staniczenko, Jason~C. Kopp, and Stefano Allesina.
\newblock The ghost of nestedness in ecological networks.
\newblock \emph{Nature Communications}, 4:\penalty0 1391, 2013.
\newblock \doi{https://doi.org/10.1038/ncomms2422}.
\newblock URL \url{https://www.nature.com/articles/ncomms2422}.

\bibitem[Vil{\`a} et~al.(2011)Vil{\`a}, Espinar, Hejda, Hulme,
  Jaro{\v{s}}{\'\i}k, Maron, Pergl, Schaffner, Sun, and
  Py{\v{s}}ek]{vila2011ecological}
Montserrat Vil{\`a}, Jos{\'e}~L Espinar, Martin Hejda, Philip~E Hulme,
  Vojt{\v{e}}ch Jaro{\v{s}}{\'\i}k, John~L Maron, Jan Pergl, Urs Schaffner, Yan
  Sun, and Petr Py{\v{s}}ek.
\newblock Ecological impacts of invasive alien plants: a meta-analysis of their
  effects on species, communities and ecosystems.
\newblock \emph{Ecology letters}, 14\penalty0 (7):\penalty0 702--708, 2011.

\bibitem[Biella et~al.(2020)Biella, Akter, Ollerton, Nielsen, and
  Klecka]{biella2020empirical}
Paolo Biella, Asma Akter, Jeff Ollerton, Anders Nielsen, and Jan Klecka.
\newblock An empirical attack tolerance test alters the structure and species
  richness of plant–pollinator networks.
\newblock \emph{Functional Ecology}, 34\penalty0 (11):\penalty0 2246--2258,
  2020.
\newblock \doi{https://doi.org/10.1111/1365-2435.13642}.
\newblock URL
  \url{https://besjournals.onlinelibrary.wiley.com/doi/abs/10.1111/1365-2435.13642}.

\bibitem[LaBar et~al.(2013)LaBar, Campbell, Yang, Albert, and
  Shea]{labar2013global}
Thomas LaBar, Colin Campbell, Suann Yang, R{\'e}ka Albert, and Katriona Shea.
\newblock Global versus local extinction in a network model of
  plant--pollinator communities.
\newblock \emph{Theoretical Ecology}, 6\penalty0 (4):\penalty0 495--503, 2013.

\bibitem[LaBar et~al.(2014)LaBar, Campbell, Yang, Albert, and
  Shea]{labar2014restoration}
Thomas LaBar, Colin Campbell, Suann Yang, R{\'e}ka Albert, and Katriona Shea.
\newblock Restoration of plant--pollinator interaction networks via species
  translocation.
\newblock \emph{Theoretical ecology}, 7\penalty0 (2):\penalty0 209--220, 2014.

\bibitem[Campbell et~al.(2015)Campbell, Yang, Albert, and
  Shea]{campbell2015plant}
Colin Campbell, Suann Yang, R{\'e}ka Albert, and Katriona Shea.
\newblock Plant--pollinator community network response to species invasion
  depends on both invader and community characteristics.
\newblock \emph{Oikos}, 124\penalty0 (4):\penalty0 406--413, 2015.

\bibitem[Campbell et~al.(2022{\natexlab{b}})Campbell, Russo, Albert, Buckling,
  and Shea]{campbell2022whole}
Colin Campbell, Laura Russo, R{\'e}ka Albert, Angus Buckling, and Katriona
  Shea.
\newblock Whole community invasions and the integration of novel ecosystems.
\newblock \emph{PLOS Computational Biology}, 18\penalty0 (6):\penalty0
  e1010151, 2022{\natexlab{b}}.

\bibitem[Fatemi~Nasrollahi et~al.(2023)Fatemi~Nasrollahi, Campbell, and
  Albert]{fatemi2023predicting}
Fatemeh~Sadat Fatemi~Nasrollahi, Colin Campbell, and R{\'e}ka Albert.
\newblock Predicting cascading extinctions and efficient restoration strategies
  in plant--pollinator networks via generalized positive feedback loops.
\newblock \emph{Scientific Reports}, 13\penalty0 (1):\penalty0 902, 2023.

\bibitem[Zhang et~al.(2008)Zhang, Shah, Yang, Nyland, Liu, Yun, Albert, and
  Loughran]{zhang_network_2008}
R.~Zhang, M.~V. Shah, J.~Yang, S.~B. Nyland, X.~Liu, J.~K. Yun, R.~Albert, and
  T.~P. Loughran.
\newblock Network model of survival signaling in large granular lymphocyte
  leukemia.
\newblock \emph{Proceedings of the National Academy of Sciences}, 105\penalty0
  (42):\penalty0 16308--16313, October 2008.
\newblock ISSN 0027-8424, 1091-6490.
\newblock \doi{10.1073/pnas.0806447105}.
\newblock URL \url{http://www.pnas.org/cgi/doi/10.1073/pnas.0806447105}.

\bibitem[Zañudo and Albert(2015)]{zanudo_cell_2015}
Jorge G.~T. Zañudo and Réka Albert.
\newblock Cell {Fate} {Reprogramming} by {Control} of {Intracellular} {Network}
  {Dynamics}.
\newblock \emph{PLOS Computational Biology}, 11\penalty0 (4):\penalty0
  e1004193, April 2015.
\newblock ISSN 1553-7358.
\newblock \doi{10.1371/journal.pcbi.1004193}.
\newblock URL \url{https://dx.plos.org/10.1371/journal.pcbi.1004193}.

\bibitem[Saadatpour et~al.(2011)Saadatpour, Wang, Liao, Liu, Loughran, Albert,
  and Albert]{saadatpour_dynamical_2011}
Assieh Saadatpour, Rui-Sheng Wang, Aijun Liao, Xin Liu, Thomas~P. Loughran,
  István Albert, and R{\'e}ka Albert.
\newblock Dynamical and structural analysis of a t cell survival network
  identifies novel candidate therapeutic targets for large granular lymphocyte
  leukemia.
\newblock \emph{PLOS Computational Biology}, 7\penalty0 (11):\penalty0 1--15,
  11 2011.
\newblock \doi{10.1371/journal.pcbi.1002267}.
\newblock URL \url{https://doi.org/10.1371/journal.pcbi.1002267}.

\bibitem[Steinway et~al.(2014)Steinway, Zañudo, Ding, Rountree, Feith,
  Loughran, and Albert]{steinway_network_2014}
Steven~Nathaniel Steinway, Jorge G.~T. Zañudo, Wei Ding, Carl~Bart Rountree,
  David~J. Feith, Thomas~P. Loughran, and Reka Albert.
\newblock Network {Modeling} of {TGF$\beta$} {Signaling} in {Hepatocellular}
  {Carcinoma} {Epithelial}-to-{Mesenchymal} {Transition} {Reveals} {Joint}
  {Sonic} {Hedgehog} and {Wnt} {Pathway} {Activation}.
\newblock \emph{Cancer Research}, 74\penalty0 (21):\penalty0 5963--5977,
  November 2014.

\bibitem[Albert et~al.(2008)Albert, Thakar, Li, Zhang, and
  Albert]{albert_boolean_2008}
István Albert, Juilee Thakar, Song Li, Ranran Zhang, and Réka Albert.
\newblock Boolean network simulations for life scientists.
\newblock \emph{Source Code for Biology and Medicine}, 3\penalty0 (1):\penalty0
  16, November 2008.
\newblock ISSN 1751-0473.
\newblock \doi{10.1186/1751-0473-3-16}.
\newblock URL \url{https://doi.org/10.1186/1751-0473-3-16}.

\bibitem[Park et~al.(2023)Park, Costa, Rocha, Albert, and
  Rozum]{park_robustness_2023}
Kyu~Hyong Park, Felipe~Xavier Costa, Luis~M. Rocha, Réka Albert, and Jordan~C.
  Rozum.
\newblock Robustness of biomolecular networks suggests functional modules far
  from the edge of chaos, July 2023.
\newblock URL
  \url{https://www.biorxiv.org/content/10.1101/2023.06.30.547297v1}.
\newblock Pages: 2023.06.30.547297 Section: New Results.

\bibitem[Klarner et~al.(2016)Klarner, Streck, and
  Siebert]{klarner_pyboolnet_2016}
Hannes Klarner, Adam Streck, and Heike Siebert.
\newblock {PyBoolNet}: a python package for the generation, analysis and
  visualization of boolean networks.
\newblock \emph{Bioinformatics}, page btw682, October 2016.
\newblock ISSN 1367-4803, 1460-2059.
\newblock \doi{10.1093/bioinformatics/btw682}.
\newblock URL
  \url{https://academic.oup.com/bioinformatics/article-lookup/doi/10.1093/bioinformatics/btw682}.

\bibitem[Trinh et~al.(2022{\natexlab{a}})Trinh, Benhamou, Hiraishi, and
  Soliman]{petre_minimal_2022}
Van-Giang Trinh, Belaid Benhamou, Kunihiko Hiraishi, and Sylvain Soliman.
\newblock Minimal {Trap} {Spaces} of {Logical} {Models} are {Maximal} {Siphons}
  of {Their} {Petri} {Net} {Encoding}.
\newblock In Ion Petre and Andrei Păun, editors, \emph{Computational {Methods}
  in {Systems} {Biology}}, volume 13447, pages 158--176, Bucarest, Romania,
  2022{\natexlab{a}}. Springer International Publishing.
\newblock ISBN 978-3-031-15033-3 978-3-031-15034-0.
\newblock \doi{10.1007/978-3-031-15034-0_8}.
\newblock URL \url{https://link.springer.com/10.1007/978-3-031-15034-0_8}.
\newblock Series Title: Lecture Notes in Computer Science.

\bibitem[Beneš et~al.(2022)Beneš, Brim, Huvar, Pastva, Šafránek, and
  Šmijáková]{benes_aeonpy_2022}
Nikola Beneš, Luboš Brim, Ondřej Huvar, Samuel Pastva, David Šafránek, and
  Eva Šmijáková.
\newblock {AEON}.py: {Python} library for attractor analysis in asynchronous
  {Boolean} networks.
\newblock \emph{Bioinformatics}, 38\penalty0 (21):\penalty0 4978--4980,
  November 2022.
\newblock ISSN 1367-4803.
\newblock \doi{10.1093/bioinformatics/btac624}.
\newblock URL \url{https://doi.org/10.1093/bioinformatics/btac624}.

\bibitem[Trinh et~al.(2022{\natexlab{b}})Trinh, Hiraishi, and
  Benhamou]{trinh_computing_2022}
Van-Giang Trinh, Kunihiko Hiraishi, and Belaid Benhamou.
\newblock Computing attractors of large-scale asynchronous boolean networks
  using minimal trap spaces.
\newblock In \emph{Proceedings of the 13th {ACM} {International} {Conference}
  on {Bioinformatics}, {Computational} {Biology} and {Health} {Informatics}},
  pages 1--10, Northbrook Illinois, August 2022{\natexlab{b}}. ACM.
\newblock ISBN 978-1-4503-9386-7.
\newblock \doi{10.1145/3535508.3545520}.
\newblock URL \url{https://dl.acm.org/doi/10.1145/3535508.3545520}.

\bibitem[Albert and Othmer(2003)]{albert_topology_2003}
Réka Albert and Hans~G Othmer.
\newblock The topology of the regulatory interactions predicts the expression
  pattern of the segment polarity genes in {Drosophila} melanogaster.
\newblock \emph{Journal of Theoretical Biology}, 223\penalty0 (1):\penalty0
  1--18, July 2003.
\newblock ISSN 00225193.
\newblock \doi{10.1016/S0022-5193(03)00035-3}.
\newblock URL
  \url{https://linkinghub.elsevier.com/retrieve/pii/S0022519303000353}.

\bibitem[Zañudo and Albert(2013)]{zanudo_effective_2013}
Jorge Gomez~Tejeda Zañudo and Réka Albert.
\newblock An effective network reduction approach to find the dynamical
  repertoire of discrete dynamic networks.
\newblock \emph{Chaos: An Interdisciplinary Journal of Nonlinear Science},
  23\penalty0 (2):\penalty0 025111, June 2013.
\newblock ISSN 1054-1500.
\newblock \doi{10.1063/1.4809777}.
\newblock URL \url{https://aip.scitation.org/doi/abs/10.1063/1.4809777}.
\newblock Publisher: American Institute of Physics.

\bibitem[Rozum et~al.(2021)Rozum, Zañudo, Gan, Deritei, and
  Albert]{rozum_parity_2021}
Jordan~C. Rozum, Jorge Gómez~Tejeda Zañudo, Xiao Gan, Dávid Deritei, and
  Réka Albert.
\newblock Parity and time reversal elucidate both decision-making in empirical
  models and attractor scaling in critical {Boolean} networks.
\newblock \emph{Science Advances}, 7\penalty0 (29):\penalty0 eabf8124, July
  2021.
\newblock ISSN 2375-2548.
\newblock \doi{10.1126/sciadv.abf8124}.
\newblock URL \url{https://advances.sciencemag.org/content/7/29/eabf8124}.
\newblock Publisher: American Association for the Advancement of Science
  Section: Research Article.

\bibitem[Yang et~al.(2018)Yang, Gómez Tejeda~Zañudo, and
  Albert]{yang_target_2018}
Gang Yang, Jorge Gómez Tejeda~Zañudo, and Réka Albert.
\newblock Target {Control} in {Logical} {Models} {Using} the {Domain} of
  {Influence} of {Nodes}.
\newblock \emph{Frontiers in Physiology}, 9, 2018.
\newblock ISSN 1664-042X.
\newblock \doi{10.3389/fphys.2018.00454}.
\newblock URL
  \url{https://www.frontiersin.org/articles/10.3389/fphys.2018.00454/full}.

\bibitem[Deritei et~al.(2016)Deritei, Aird, Ercsey-Ravasz, and
  Regan]{deritei_principles_2016}
Dávid Deritei, William~C. Aird, Mária Ercsey-Ravasz, and Erzsébet~Ravasz
  Regan.
\newblock Principles of dynamical modularity in biological regulatory networks.
\newblock \emph{Scientific Reports}, 6\penalty0 (1):\penalty0 21957, April
  2016.
\newblock ISSN 2045-2322.
\newblock \doi{10.1038/srep21957}.
\newblock URL \url{http://www.nature.com/articles/srep21957}.

\bibitem[Deritei et~al.(2019)Deritei, Rozum, Regan, and
  Albert]{deritei_feedback_2019}
D{\'a}vid Deritei, Jordan~C Rozum, Erzs{\'e}bet~Ravasz Regan, and R\'eka
  Albert.
\newblock A feedback loop of conditionally stable circuits drives the cell
  cycle from checkpoint to checkpoint.
\newblock \emph{Scientific Reports}, 9:\penalty0 16430, 2019.
\newblock \doi{10.1038/s41598-019-52725-1}.

\bibitem[Paulevé et~al.(2020)Paulevé, Kolčák, Chatain, and
  Haar]{pauleve_reconciling_2020}
Loïc Paulevé, Juraj Kolčák, Thomas Chatain, and Stefan Haar.
\newblock Reconciling qualitative, abstract, and scalable modeling of
  biological networks.
\newblock \emph{Nature Communications}, 11\penalty0 (1):\penalty0 4256, August
  2020.
\newblock ISSN 2041-1723.
\newblock \doi{10.1038/s41467-020-18112-5}.
\newblock URL \url{https://www.nature.com/articles/s41467-020-18112-5}.

\bibitem[Nasrollahi et~al.(2021)Nasrollahi, Za{\~n}udo, Campbell, and
  Albert]{nasrollahi2021relationships}
Fatemeh Sadat~Fatemi Nasrollahi, Jorge G{\'o}mez~Tejeda Za{\~n}udo, Colin
  Campbell, and R{\'e}ka Albert.
\newblock Relationships among generalized positive feedback loops determine
  possible community outcomes in plant-pollinator interaction networks.
\newblock \emph{Physical Review E}, 104\penalty0 (5):\penalty0 054304, 2021.

\bibitem[Zañudo et~al.(2017)Zañudo, Yang, and
  Albert]{zanudo_structure-based_2017}
Jorge Gomez~Tejeda Zañudo, Gang Yang, and Réka Albert.
\newblock Structure-based control of complex networks with nonlinear dynamics.
\newblock \emph{Proceedings of the National Academy of Sciences}, 114\penalty0
  (28):\penalty0 7234--7239, July 2017.
\newblock ISSN 0027-8424, 1091-6490.
\newblock \doi{10.1073/pnas.1617387114}.
\newblock URL \url{https://www.pnas.org/content/114/28/7234}.
\newblock Publisher: National Academy of Sciences Section: Physical Sciences.

\bibitem[Fiedler et~al.(2013)Fiedler, Mochizuki, Kurosawa, and
  Saito]{fiedler_dynamics_2013}
Bernold Fiedler, Atsushi Mochizuki, Gen Kurosawa, and Daisuke Saito.
\newblock Dynamics and {Control} at {Feedback} {Vertex} {Sets}. {I}:
  {Informative} and {Determining} {Nodes} in {Regulatory} {Networks}.
\newblock \emph{Journal of Dynamics and Differential Equations}, 25\penalty0
  (3):\penalty0 563--604, September 2013.
\newblock ISSN 1040-7294, 1572-9222.
\newblock \doi{10.1007/s10884-013-9312-7}.
\newblock URL \url{http://link.springer.com/10.1007/s10884-013-9312-7}.

\bibitem[Mochizuki et~al.(2013)Mochizuki, Fiedler, Kurosawa, and
  Saito]{mochizuki_dynamics_2013}
Atsushi Mochizuki, Bernold Fiedler, Gen Kurosawa, and Daisuke Saito.
\newblock Dynamics and control at feedback vertex sets. {II}: {A} faithful
  monitor to determine the diversity of molecular activities in regulatory
  networks.
\newblock \emph{Journal of Theoretical Biology}, 335:\penalty0 130--146,
  October 2013.
\newblock ISSN 0022-5193.
\newblock \doi{10.1016/j.jtbi.2013.06.009}.
\newblock URL
  \url{https://www.sciencedirect.com/science/article/pii/S0022519313002816}.

\bibitem[Snoussi and Thomas(1993)]{snoussi_logical_1993}
El~Houssine Snoussi and Ren\'e Thomas.
\newblock Logical identification of all steady states: {The} concept of
  feedback loop characteristic states.
\newblock \emph{Bulletin of Mathematical Biology}, 55\penalty0 (5):\penalty0
  973--991, 1993.

\bibitem[Richard(2019)]{richard_cycles_2019}
Adrien Richard.
\newblock Positive and negative cycles in boolean networks.
\newblock \emph{Journal of Theoretical Biology}, 463:\penalty0 67--76, 2019.
\newblock ISSN 0022-5193.
\newblock \doi{https://doi.org/10.1016/j.jtbi.2018.11.028}.
\newblock URL
  \url{https://www.sciencedirect.com/science/article/pii/S0022519318305812}.

\bibitem[Newby et~al.(2022)Newby, Tejeda~Za{\~{n}}udo, and
  Albert]{newby-chaos-2022}
Eli Newby, Jorge~G{\'o}mez Tejeda~Za{\~{n}}udo, and R{\'e}ka Albert.
\newblock Structure-based approach to identifying small sets of driver nodes in
  biological networks.
\newblock \emph{Chaos: An Interdisciplinary Journal of Nonlinear Science},
  32\penalty0 (6):\penalty0 063102, 2022.
\newblock \doi{10.1063/5.0080843}.
\newblock URL \url{https://doi.org/10.1063/5.0080843}.

\bibitem[Santolini and Barab{\'a}si(2018)]{santolini_PNAS_2018}
Marc Santolini and Albert-L{\'a}szl{\'o} Barab{\'a}si.
\newblock Predicting perturbation patterns from the topology of biological
  networks.
\newblock \emph{Proceedings of the National Academy of Sciences}, 115\penalty0
  (27):\penalty0 E6375--E6383, 2018.
\newblock \doi{10.1073/pnas.1720589115}.
\newblock URL \url{https://www.pnas.org/doi/abs/10.1073/pnas.1720589115}.

\bibitem[Zhirov et~al.(2010)Zhirov, Zhirov, and Shepelyansky]{zhirov_2010}
A.~O. Zhirov, O.~V. Zhirov, and D.~L. Shepelyansky.
\newblock Two-dimensional ranking of wikipedia articles.
\newblock \emph{The European Physical Journal B}, 77\penalty0 (4):\penalty0
  523--531, Oct 2010.
\newblock ISSN 1434-6036.
\newblock \doi{10.1140/epjb/e2010-10500-7}.
\newblock URL \url{https://doi.org/10.1140/epjb/e2010-10500-7}.

\bibitem[Feo and Resende(1995)]{feo_1995}
Thomas~A. Feo and Mauricio G.~C. Resende.
\newblock Greedy randomized adaptive search procedures.
\newblock \emph{Journal of Global Optimization}, 6\penalty0 (2):\penalty0
  109--133, Mar 1995.
\newblock ISSN 1573-2916.
\newblock \doi{10.1007/BF01096763}.
\newblock URL \url{https://doi.org/10.1007/BF01096763}.

\bibitem[Steinway et~al.(2015)Steinway, Za{\~{n}}udo, Michel, Feith, Loughran,
  and Albert]{steinway_npjSystemsBio_2015}
Steven~Nathaniel Steinway, Jorge G{\'o}mez~Tejeda Za{\~{n}}udo, Paul~J. Michel,
  David~J. Feith, Thomas~P. Loughran, and R{\'e}ka Albert.
\newblock Combinatorial interventions inhibit tgf$\beta$-driven
  epithelial-to-mesenchymal transition and support hybrid cellular phenotypes.
\newblock \emph{npj Systems Biology and Applications}, 1\penalty0 (1):\penalty0
  15014, Nov 2015.
\newblock ISSN 2056-7189.
\newblock \doi{10.1038/npjsba.2015.14}.
\newblock URL \url{https://doi.org/10.1038/npjsba.2015.14}.

\bibitem[Campbell and Albert(2019)]{campbell-chaos-2019}
Colin Campbell and R{\'e}ka Albert.
\newblock Edgetic perturbations to eliminate fixed-point attractors in boolean
  regulatory networks.
\newblock \emph{Chaos: An Interdisciplinary Journal of Nonlinear Science},
  29\penalty0 (2):\penalty0 023130, 2019.
\newblock \doi{10.1063/1.5083060}.
\newblock URL \url{https://doi.org/10.1063/1.5083060}.

\bibitem[Shmulevich and Dougherty(2010)]{shmulevich_probabilistic_2010}
Ilya Shmulevich and Edward~R Dougherty.
\newblock \emph{Probabilistic Boolean Networks: The Modeling and Control of
  Gene Regulatory Networks}.
\newblock Society of Industrial and Applied Mathematics, 2010.

\end{thebibliography}

\section*{Acknowledgements}

This work was supported by National Science Foundation grants IIS 1814405 and MCB 1715826 and Army Research Office grant 79961-MS-MUR.

\end{document}